\documentclass{article}

\usepackage{arxiv}
\usepackage{amsmath}

\usepackage{caption} 
\usepackage{subcaption} 

\usepackage[utf8]{inputenc} 
\usepackage[T1]{fontenc}    
\usepackage{hyperref}       
\usepackage{url}            
\usepackage{booktabs}       
\usepackage{amsfonts}       
\usepackage{nicefrac}       
\usepackage{microtype}      
\usepackage{lipsum}		
\usepackage{graphicx}
\usepackage{natbib}
\usepackage{doi}

\title{From Catastrophic to Concrete: Reframing AI Risk Communication for Public Mobilization}

\author{{Philip Trippenbach} \\
	Seismic Foundation\\
	\texttt{philip@seismic.org} \\
	\And
	{Isabella Scala} \\
	Seismic Foundation\\
	\texttt{isabella@seismic.org} \\
	 \AND
	 Jai Bhambra \\
	 Blue State \\
	 \texttt{jbhambra@bluestate.co} \\
	 \And
	 Rowan Emslie \\
	 Centre for Future Generations \\
	 \texttt{r.emslie@cfg.eu} \\
}

\date{November 9, 2025}

\renewcommand{\shorttitle}{From Catastrophic to Concrete: Reframing AI Risk Communication for Public Mobilization}

\hypersetup{
pdftitle={A template for the arxiv style},
pdfsubject={q-bio.NC, q-bio.QM},
pdfauthor={David S.~Hippocampus, Elias D.~Striatum},
pdfkeywords={First keyword, Second keyword, More},
}

\begin{document}
\maketitle

\begin{abstract}
Effective governance of artificial intelligence (AI) requires public engagement, yet communication strategies centered on existential risk have not produced sustained mobilization. In this paper, we examine the psychological and opinion barriers that limit engagement with extinction narratives, such as mortality avoidance, exponential growth bias, and the absence of self-referential anchors. We contrast them with evidence that public concern over AI rises when framed in terms of proximate harms such as employment disruption, relational instability, and mental health issues. We validate these findings through actual message testing with 1063 respondents, with the evidence showing that AI risks to Jobs and Children have the highest potential to mobilize people, while Existential Risk is the lowest-performing theme across all demographics. Using survey data from five countries, we identify two segments (Tech-Positive Urbanites and World Guardians) as particularly receptive to such framing and more likely to participate in civic action. Finally, we argue that mobilization around these everyday concerns can raise the political salience of AI, creating “policy demand” for structural measures to mitigate AI risks. We conclude that this strategy creates the conditions for successful regulatory change.
\end{abstract}


\section{Introduction}

\textit{Should we attempt to engage the public in the risks of AI? The evidence shows we should.}

AI is a powerful technology, and, like all powerful technologies, comes with risks. These risks include the possibility of catastrophic damage, loss of control, and human extinction. (\citealt{8}). We need sensible policies in place to mitigate these risks (\citealt{10}; \citealt{8}). Policy is not created in a vacuum; in democratic countries, policymakers are responsive to public support and demand, at least in critical constituencies. The relationship between policy and public support and interest is complex, but well-established and intensively studied (\citealt{66}; \citealt{15}; \citealt{72}).

Furthermore, the importance of an issue in policy formation is connected to its \textit{salience}, or in other words, its ranking by the public among problems perceived as important (\citealt{48}). Though much research shows that public opinion is broadly in favour of strong regulations on AI (\citealt{14}), the issue nevertheless ranks low when respondents are asked to rank it alongside other issues of public interest (\citealt{57}). When an issue is \textit{not} of high public salience, elite and institutional factors, rather than a popular mandate, are the primary determinants of policy outcomes (\citealt{6}). In the current environment, the power of the tech lobby and the disposition of the Trump administration to tech policy suggest that, absent large-scale public mobilization, default policy outcomes are likely to be pro-tech and pro-acceleration rather than focused on risk mitigation, as we have seen with America’s AI Action Plan (\citealt{41}; Pearl, 2025). This is backed by the historical example of the political success of the Crypto lobby in the USA, activity now being echoed in the establishment of several Super PACs in the USA, intended to help elect legislators inclined to avoid AI regulation (\citealt{34}). 

Conversely, historical evidence suggests that mobilizing the public on a matter of public interest—i.e., raising its salience—can be effective in altering policy. There is widespread empirical evidence that public issue salience campaigns from a wide array of social movements and interest groups have been effective in putting issues onto the political agenda, from civil rights (\citealt{65}) to climate (\citealt{12}) to gun rights (\citealt{43}). Furthermore, the connection between interest groups' activities and actual policy change is also well established, with nearly 300 specific interest groups credited with a measurable policy outcome since 1945 in the US alone (\citealt{32}).

Media and communications activity can be a successful way of doing this. Fictional films such as \textit{The Day After} in 1983 (\citealt{55}), and documentaries such as \textit{An Inconvenient Truth} in 2006 (\citealt{45}; \citealt{7}), \textit{Blackfish} in 2013 (\citealt{11}), \textit{Cowspiracy} in 2014 (\citealt{47}), and \textit{13th} in 2016 (\citealt{22} have proven successful in raising the salience of an issue and mobilizing the public. More recently, evidence suggests that public discussion of technology is influenced by its presentation in popular media such as the \textit{Black Mirror} video franchise (\citealt{49}).

\textbf{Therefore, we need to mobilize at least some groups of the public to demand action, with a view to building public support for adequate risk-mitigation policies for AI.} 

\textit{How do we mobilize the public?}

\section{Barriers to Communication on Existential Risk}
\textit{Should we talk to people directly about the most severe, existential risks of AI? The evidence suggests that this approach will not be effective in building the required civil society movement.}

\begin{figure}[htbp]
    \centering
    \begin{subfigure}[b]{0.8\textwidth}
        \includegraphics[width=\textwidth]{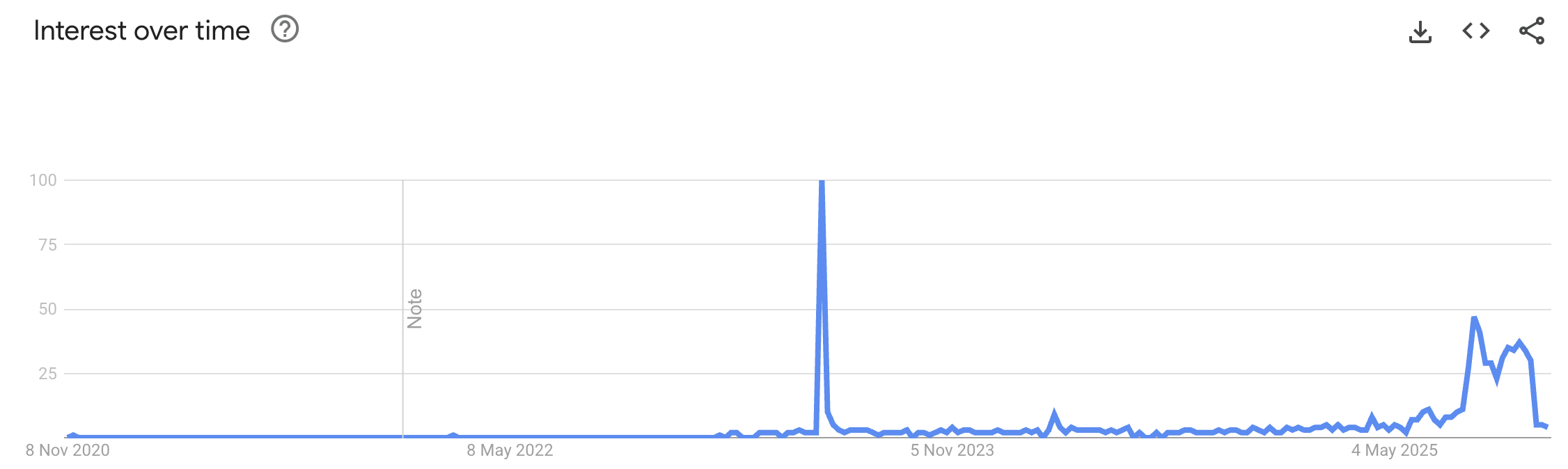}
        \caption{Search interest in “AI extinction” shows a growing interest, albeit from a very low base. The growth is also uneven. The peak in late May 2023 represents the release of the CAIS statement on AI. We can see here that major events in the AI safety movement, such the Bletchley Park summit (November 2023), had little effect on public interest in AI extinction. The smaller peak towards the end of the graph, in mid-August 2025, may seem important, but on closer inspection, it is incidental to a larger story - see below. (Global trends, Nov 2021-Oct 2025)}
        \label{fig:1a}
    \end{subfigure}
    \vfill 
    \begin{subfigure}[b]{0.8\textwidth}
        \includegraphics[width=\textwidth]{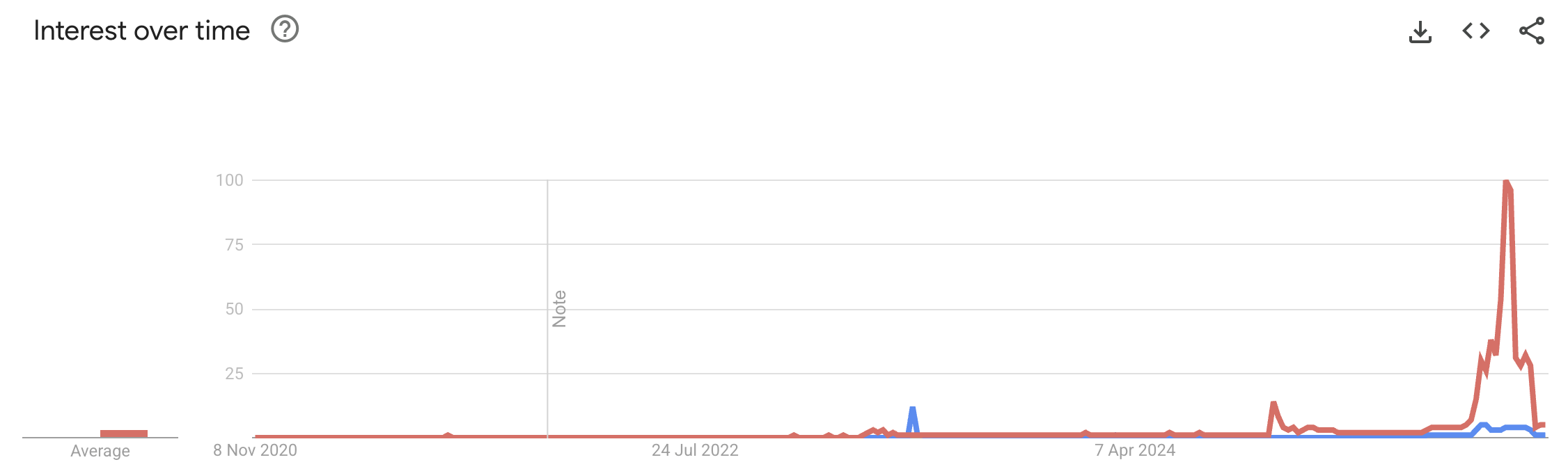}
        \caption{Here is the same data, but comparing searches for “AI extinction” (blue line) with searches for “AI suicide” (red line). The peaks on the red line correspond to news coverage of the suicides of an anonymous man in Belgium (April 2023), Sewell Setzer III, a 14-year-old from Florida (October 2024), and Adam Reine, a 16-year-old from California (August 2025). Here we can see that the peak in “AI extinction” searches in August of 2025 was connected to a large and sudden increase in searches for AI suicide and not an independent event. (Global trends, Nov 2021-Oct 2025)}
        \label{fig:1b}
    \end{subfigure}
    \vfill 
    \begin{subfigure}[b]{0.8\textwidth}
        \includegraphics[width=\textwidth]{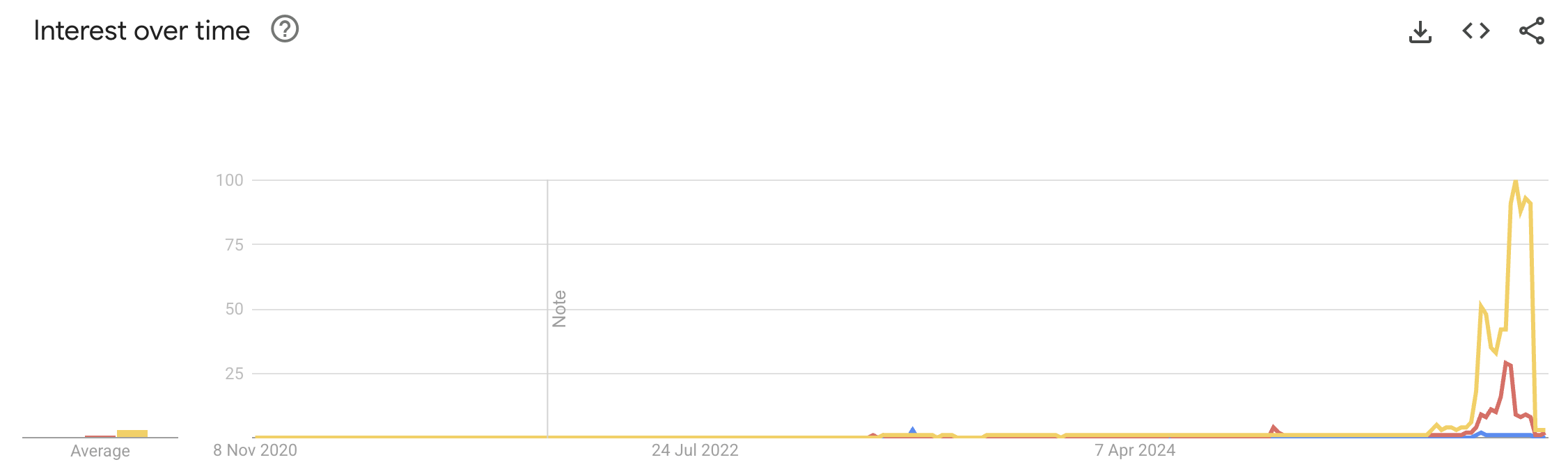}
        \caption{And, finally, here is the same data but comparing searches for “AI extinction” (blue line) with searches for “AI suicide” (red line) and “AI regulation” (yellow line). Here, we can clearly see that past AI safety events have not been successful in driving interest in AI regulation, when compared to the current interest in AI suicide. (Global trends, Nov 2021-Oct 2025)}
        \label{fig:1c}
    \end{subfigure}
    \caption{Google Trends search data shows the relative public interest in various aspects of AI. }
    \label{fig:1}
\end{figure}

Much effort and attention has been focused on communicating to the public on the risks of AI, up to and including the existential risk of AI causing human extinction. The much-publicized CAIS statement on AI Risk (\citealt{17}) appears, to date, to be the most successful of these efforts in terms of capturing public and media attention. This was not an isolated action; since the launch of ChatGPT, the world’s attention has been focused on AI, with much prominent news coverage devoted to statements by authority figures warning of AI risks (see, e.g., \citealt{61}; \citealt{9}). However, there is little strong evidence that this has led to increased public interest in AI regulation (See Fig. \ref{fig:1a}). 

Why have these high-profile calls not resulted in public mobilization? 

We contend that attempting to engage people directly on the existential risk of AI may not be an optimal strategy, or even a viable one. 

There are strong reasons for this, grounded in psychological science and public opinion research. Let us begin with the psychological factors.

\textbf{Psychological Barriers}

People simply don’t react well to being told about the existential risk of AI.\footnote{This has been a subject of discussion within the AI safety community for some time; see e.g. (\citealt{67}).}  There are three reasons for this, grounded in well-established principles of human psychology. They are, first, the universal human tendency to avoid reminders of mortality; second, the tendency to misapprehend the nature of exponential growth, and third, the tendency to have difficulty engaging with information that lacks personal reference points. 

\textbf{Mortality Avoidance Behavior}

The first reason is unwillingness to engage with discussions of one’s own mortality. 

People tend to avoid messaging about mortality, as awareness of death generates profound anxiety. Research shows that, when confronted with reminders of their mortality, individuals tend to actively defend against this through sophisticated psychological mechanisms, collectively known as Terror Management Theory (TMT) (\citealt{30}; \citealt{31}) 

In short, when mortality is made salient, people respond with heightened defensiveness to protect their self-esteem and pre-existing cultural worldview. This defensive response leads individuals to minimize, avoid, or reject mortality-related information, which explains why death-related messaging often fails to achieve engagement despite its universal personal relevance (\citealt{53}). The psychological literature is replete with examples of people’s inherent reluctance to engage in conversations that increase the salience of mortality, in multiple contexts, even when engaging with trusted loved ones (e.g., cancer care in \citealt{4})

This effect has recently been demonstrated to be particularly applicable to advertising messaging. Increasing the salience of mortality has been shown to drive people to resist change and cling harder to a pre-existing worldview (\citealt{38}). 

The stark nature of a message as bold as “if anyone builds it, everyone dies” may therefore be ineffective, as it directly triggers most people's inherent psychological defense mechanisms. TMT shows us that, after catching public attention due to its boldness, messaging like this is likely to be actively dismissed by most members of the public.\footnote{This conclusion is no doubt counterintuitive to readers who are members of the AI safety community, and therefore by definition already concerned and mobilized on the topic of AI’s existential risks. \textit{We} were all moved to action by the early warnings of mortal danger; how could it be otherwise for the majority of the population? The evidence suggests that the members of the existing AI safety community are unusual in their ability to engage with messages that increase the salience of mortality while maintaining effectiveness. This may be an interesting question for further research. Regardless, the conclusion is the same: it is not sufficient merely to spread the message of existential risk wider, as much of the population may be constitutionally incapable of engaging with it.}

\textbf{Exponential Growth Bias}

Another reason people have difficulty accepting and engaging with existential risk messaging is humans’ almost universally poor appreciation of exponential growth. 

\textit{Exponential growth bias} refers to the phenomenon in which humans intuitively underestimate exponential growth. This facet of psychology has been well-known for some time and even immortalized in the story of the maharajah and the chessboard (\citealt{42}). Modern experimental results have confirmed how widespread this bias is (\citealt{63}), and the phenomenon has been shown to apply even to highly educated people, \textit{even if they are aware of exponential growth bias} (\citealt{56}). 

This specific bias is important because it is critical in appreciating the urgency of the need to regulate AI. This is because the exponential and accelerating potential of recursive self-improvement needs to be properly appreciated to understand its compounding effect on shortening timelines to transformational AI. 

With a poor understanding of exponential growth, the effectiveness of media such as \textit{AI 2027} (\citealt{37}) is reduced. Though content like this may gain short-term media attention (as AI 2027 did), that does not automatically translate to public engagement and action; the public lacks an intuitive understanding of the exponential growth it describes, so the scenario ‘feels’ unrealistic, and it is therefore more easily dismissed.

\textbf{Self-Reference Effect} 

A third reason is that the dangers of runaway AI are abstract. Research shows that people do not react well or engage easily with messaging around issues they have no personal experience with, or that they cannot easily relate to themselves and their personal experiences (\citealt{52}; \citealt{60}).

For AI risk messaging, this psychological feature of humans is problematic because the risks of an AI takeover scenario are unprecedented and quite abstract. There is no precedent in human experience for being disassembled at a molecular level by swarms of nanomachines, or being outcompeted as superintelligent AI builds incomprehensible megaprojects, especially as described in fast take-off scenarios (e.g., \citealt{37}; \citealt{27})

The actual development of technology and our responses to it are influenced by popular culture (\citealt{21}). Indeed, popular culture and mainstream science fiction have provided the public with familiar reference points for imagining loss-of-control scenarios, from The \textit{Matrix} and \textit{Terminator} to \textit{Westworld} and \textit{Ex Machina}. These stories help audiences understand catastrophic risks, but they may not be as effective as they often rely on Hollywood spectacle. While this dramatization highlights the moral urgency of the issue, it also fosters a distorted understanding of actual capabilities and dangers (\citealt{5}; \citealt{28}). The risk is that serious debate gets blurred with blockbuster tropes – charismatic heroes, killer robots, machine apocalypse – leaving audiences with the impression that AI takeover belongs more to the realm of superhero villains than sober, policy-relevant debate.

The Self-Reference Effect poses a challenge for messaging about the existential risks of AI because of the unprecedented nature of AI risks. Since these risks have never been encountered before in human history, let alone individual experience, \textit{accepting them feels like a leap of faith} to most people. This genuine psychological reality applies \textit{despite} the fact that the existential risks of AI are borne out through logical analysis and supported by leading experts in the field of AI research. 

The rationality of the argument in favor of AI risk simply fails in the face of the genuine lived experience of most people.

\textbf{Public Opinion Barriers}

Public opinion research confirms the effect of the psychological barriers described above. AI has been prominent in the media and public culture for over a decade (\citealt{35}). Since the launch of ChatGPT in November of 2022, AI has shot to the top of public imagination and media attention. And, nevertheless, three years of prominence for the message of AI’s existential risk have not resulted in mass public mobilization.

This is despite the fact that polling shows concern for loss-of-control scenarios with AI holding steady, with around 60\% of the population saying they are concerned in various polls (e.g., \citealt{23}; \citealt{57}). Other studies present a mixed picture, with some polls suggesting a decline in concern about AI harms (\citealt{70}) and others showing cautious optimism growing about AI (\citealt{16}). While some polls show concerns about \textit{specific} AI harms are growing (e.g. job loss, deepfakes), there is no strong evidence for growing public concern about widespread existential risks. 

Indeed, a growing body of public opinion research confirms that the catastrophic risks of AI are considered unlikely by both experts and the general public (\citealt{33}; \citealt{68}; \citealt{57}). 

Data on search trends corroborates this information (see Fig. \ref{fig:1}). Despite the apparent success of the Center for AI Safety Statement on AI Risk, the Future of Life Institute (FLI) Pause letter, and other landmark international events such as the Bletchley Park summit, public searches for “AI extinction” have remained extremely low. Searches for “AI regulation” have been near zero throughout, until the recent prominence of AI suicide stories in the press in late August 2025. 

\textbf{Therefore, we need to find ways of mobilizing the public on AI, but using themes and narratives they can easily relate to and engage with.}

\textit{So how do we reach them?}

\section{Communication Opportunities}
\textit{What themes and narratives should we use to engage people on the risks of AI, and raise the issue’s salience in civil society? A variety of opportunities exist to make the topic emotionally relevant to the public.}

“Much like LLMs, lots of people don’t really have world models. They believe what their friends believe, or what has good epistemic vibes,” in the words of Scott Alexander.

There is an extensive and longstanding body of information and media theory research showcasing Mr. Alexander's point. The early breakthrough came in the 1940s, as scholars investigated the impact of mass media on electoral outcomes, finding that in-person interactions with opinion leaders within communities were the key source of influence on voter decisionmaking (\citealt{39}; \citealt{36}). This observation has been repeated in multiple countries (\citealt{58}; \citealt{46}) and has been updated for the modern digital media landscape (\citealt{19}), where the role of policy influencer has become a widespread and often online phenomenon (\citealt{59}).

This social coding of influence and its effect on decision-making may be one reason why the logic-based arguments around existential risk have failed to ignite the public imagination. People are emotional decision-makers (\citealt{71}). To mobilize people and change their behaviour, we need to speak to them on matters relevant to their interests, that is, socially positive (or at least acceptable) and in a manner suited to mobilizing them. 

We have seen how some factors of human psychology work against accepting catastrophic risk narratives; let us now explore how these same factors can be put to use. 

While public engagement with existential risk narratives has been low, the public has reacted to stories of AI-driven suicide with a high level of interest (see Fig. \ref{fig:1}). This illustrates some critical components of psychology that apply to communicating about AI risk.

\textbf{The Availability Heuristic}

The first is the availability heuristic. This phenomenon occurs when people judge the likelihood or frequency of events based on how easily examples come to mind, rather than on actual statistical probability (\citealt{62}). So, in the tragic case of Adam Raine’s death, the availability heuristic leads people to believe this type of outcome is more common than it may be in reality. Stories of demonstrated, present-day harms that vividly illustrate the risks of AI – regardless of how common they may be – are therefore relevant to raising the salience of AI as a matter of public policy. 

\textbf{The Self-Reference Effect}

In itself, availability is not enough. People are also more likely to engage with information that they feel affects them directly and personally. This is formally known as the Self-Reference Effect. This is the tendency for people to remember information better when it has been encoded in reference to the self, i.e., information that can be directly related to personal experience (\citealt{52}; \citealt{60}).

The death of a child by suicide is among the most tragic possible events that may occur in a life, and yet it is easy to imagine, especially for a parent. Moreover, every teenager and young adult can relate to the feelings of alienation they had in their youth. Meanwhile, to a parent, the grief of losing a child weighs so heavily that even its mere future potential casts a shivering echo into the present. This is why the search data show us an extraordinary response to AI suicide stories from the public, and a concomitant increase in searches on AI regulation (see Fig. \ref{fig:1}). 

Stories of AI bias, or job disruption, or, yes, even human extinction, may seem abstract and far removed from a person’s daily life. But the idea that a child could be so taken by their conversation with an AI that it led them to take their own life strikes an emotional note of rare, shrill purity.\footnote{Eagle-eyed readers may note that the 1983 film \textit{The Day After} was successful in affecting public opinion and policy on nuclear weapons. Isn’t this also an existential threat, and an abstract one? In this specific case, we disagree. While a full nuclear war has not happened, we have seen two cities destroyed with nuclear weapons. The images of Hiroshima, Nagasaki and the aftermath are part of every schoolchild’s curriculum. More importantly, in 1983 this reality was not remote from human experience. Moreover, though relatively few people saw the devastation in those two cities, many millions more had personally experienced the devastation of the second world war, with firebombed cities laid waste across Europe and Asia. The nuclear non-proliferation movement of the 1950s and 60s, which gave us international treaties and the International Atomic Energy Agency (IAEA), was forged by people with personal memories of the second world war. This was still very much a part of lived human experience in 1983, and therefore anti-nuclear campaigners could take advantage of the Self-Reference effect at that time.}  The current public response to this story underscores this.

\textbf{Topics of Current Interest}

Therefore, since we cannot effectively engage people on existential risk due to Terror Management Theory, Exponential Growth Bias, and the Self-Reference Effect, we must use the Self-Reference Effect and Availability Heuristic to connect with people on topics that are proximate, easily relatable, and that they already care about. 

\textit{So, what do people already care about when it comes to AI?}

Our own research on this topic involved fieldwork in five countries: the United States, the United Kingdom, France, Germany, and Poland. We surveyed approximately 2,000 people in each country, representatively sampled to reflect the country as a whole accurately. Fieldwork was conducted in June of 2025. Full methodology, sample breakdown, and segmentation information are available in the full report (see \citealt{57}). 

\textbf{AI itself is of low salience}

When the question is narrowly construed, we found that the public considers AI low on a list of important problems (\citealt{57}); unprompted members of the public rank the use of AI as a less important problem than war, climate change, mental health, unemployment, and other problems. This finding is consistent with earlier surveys showing relatively low salience for societal risks connected to AI (e.g., \citealt{68}), despite the vastly increased prominence of AI in current media and public attention. 

However, despite this low salience for AI as a topic in and of itself, we also identified areas where public concern over AI could be mobilized through its role in proximate topics that \textit{are} of interest to the public. 

Concern for AI appears to be interwoven with other narratives in the public mind. Understanding these is critical for the development of effective civil society interventions. 

First, we are confident that there is potential for mobilizing the public due to the general public attitude towards AI. The public is split on AI’s potential to benefit humanity, with only 31\% of respondents confident that it will (\citealt{57}). 

Furthermore, optimism about AI is concentrated among males and those who are better off; female respondents and people in lower income quintiles were substantially more concerned about the negative impacts of AI than respondents in general (\citealt{57}). 

\textbf{AI as part of the grievance complex}

This points us to a deeper realization about the way AI figures in the public psyche. In addition to being more optimistic about AI, wealthier respondents are also more likely to trust in existing institutions overall (\citealt{24}). This is because, from their position of socioeconomic advantage, the system appears to be working. 

In contrast, individuals from disadvantaged groups and those who are less well-off are more distrustful of institutions and much more likely to hold a sense of grievance (\citealt{24}). This sense of grievance is prevalent internationally and has been a figure of political rhetoric in many countries.

We suggest that these factors are connected. AI may be perceived as a net positive by people in a position of socioeconomic advantage because it is perceived as a facet of the system itself. 

In contrast, those who are not in positions of power are much more skeptical of AI, as they are already concerned that the system is not working in their favor.

\textbf{Put another way, AI is not perceived to be a problem in and of itself by the broad mass of the public; rather, AI is seen as a prominent and salient part of a complex of social, technological, and economic factors that are making life in the 21st century more pressured and less stable.} 

This important insight offers us a way to engage people in mitigating the risks of AI. 

We saw in section 2 that engaging people on the catastrophic risks of AI faces steep barriers of psychology and opinion. In contrast, our research reveals a considerable latent interest in AI as an aspect of current politics. We believe that critical constituencies of the public can be effectively mobilized on this issue. 

We will now turn to the topics of most salience in this approach, and then to the audience subgroups likely to be the easiest to mobilize. 

Two specific areas of relevance were identified in this research, clustered around (1) Jobs \& Employment, and (2) Children \& Relationships.

\textbf{First high-engagement theme: Jobs \& Employment}

AI’s effect on jobs was one of the most frequently mentioned concerns in our research, with 57\% of the public expressing concern about AI’s effect on jobs and employment (this effect is more pronounced in lower-income quintiles; see discussion in the previous section). The current prominence of this topic in the media has been heightened by recent studies (\citealt{13}) and statements by AI leaders (\citealt{2}; \citealt{3}). 

The issue has become one of explicit mainstream concern. Earlier this year, the European Commission’s annual Strategic Foresight report – delivered as a forward looking counterpoint to the Commission President’s State of the European Union address since 2020 – referenced AI as one of the global megatrends that presents “potential productivity increases as well as labour-market disruption” (\citealt{25}), citing EU-focused research on that topic (\citealt{18}). 

Students, especially, are concerned about AI affecting their future employment prospects. A majority of them (60\%) are concerned that they will struggle to find work in a world with AI, and 57\% of them say they feel daunted by the prospect of the future. This concern may be well-founded, as some recent studies are starting to show that entry-level work is being affected by AI displacement (\citealt{13}). 

It is critical to note that the psychological effect of this worry is connected to perceptions of personal value and dignity, rather than simply the fact of employment and earning a salary. Respondents frequently mentioned they were worried about AI affecting their employment prospects, despite unemployment in the USA being at a low 4.2\% at the time of writing. This suggests that it is not merely the availability of jobs, but also their quality, that concerns respondents. Respondents frequently discussed themes of respect and replaceability in focus group work (\citealt{57}). These findings align with broader public attitudes. A 2025 Pew Research Center survey of more than 5,200 U.S. workers found that they are worried about the future impact of AI in the workplace, and 32\% believe it will lead to fewer job opportunities for them in the long run. While 36\% report feeling hopeful, nearly as many, 33\%, feel overwhelmed about how AI will be used in the workplace. Approximately one in six workers already utilize AI in some aspect of their job, and another 25\% believe their work could be automated by AI (\citealt{50}). A Reuters/Ipsos poll similarly found that 71\% of Americans are concerned that AI will permanently put too many people out of work, and 77\% worry it could be used to stir political chaos (\citealt{51}).

European attitudes present a more optimistic contrast. A 2025 Eurobarometer summary reports that over 60\% of Europeans view AI and robotics in the workplace positively, and more than 70\% believe they improve productivity. Although 84\% insist AI must be carefully managed to protect privacy and ensure transparency in workplace decisions (\citealt{26} Academic evidence also indicates the issue is not only job quantity but also dignity, fairness, and control. Reviews of AI in HR reveal efficiency gains alongside concerns about job security, fairness, privacy, and mental health, with transparency and employee involvement emerging as key factors in building trust and satisfaction (\citealt{54}). Complementing these perception data, a longitudinal study using Germany's SOEP (Socio-Economic Panel) from 2000 to 2020 finds no sizeable impact of AI on workers' well-being or mental health, and if anything, improved in self-rated health and health satisfaction, plausibly via reduced physical job intensity, in the context of strong labor institutions (\citealt{29})

It is instructive that the European findings are taken from countries where, in general, labour protections are stronger than in the USA. Taken together, these results indicate that public concern about AI is grounded in questions of autonomy, respect, and the quality of work, and that outcomes depend on governance, transparency, and institutional context, rather than AI adoption alone.

\textbf{Second high-engagement theme: Children \& Relationships } 

Children \& Relationships stood out as another area of great concern to the polled public. In fact, polling shows that a greater proportion of the public is concerned about AI’s disruptive effect on relationships than about its disruptive effect on jobs (60\% vs. 57\%) (\citealt{57}). This is not unique to this research. The Reuters/Ipsos poll similarly found that two-thirds of Americans fear people may replace human relationships with AI companions, and a 2025 Pew Research Center survey reported that 42\% of respondents are concerned about AI weakening social bonds (\citealt{51}; \citealt{50}). Further polling reveals the depth of these concerns: a 2024 Institute for Family Studies/YouGov survey found that only one in four (25\%) young adults believe AI has the potential to replace real-life romantic relationships, while only 7\% of single young adults say they would be open to having an AI romantic partner. About 11\% are open to having an AI friend, with 1\% already doing so, and roughly one-third (32\%) remain undecided (\citealt{64}). Importantly, higher levels of personal disclosure to AI companions and reliance on them among those with smaller offline social networks were consistently associated with lower well-being (\citealt{69}). The fact that these figures match or exceed levels of concern about job displacement suggests that public anxiety about AI's social consequences is not only widespread but potentially intensifying as generative AI becomes more embedded in everyday life. 

The current level of interest in AI psychosis and AI suicides – especially the tragic recent suicides of minors – underscores just how ripe the public is for engagement on this topic. 

Parents are especially concerned, with 52\% of parents worried about their children developing a friendship with an AI and 69\% worried about their children developing a romantic relationship with an AI (\citealt{57}). Note that these figures come from polling conducted \textit{before} the news stories about Adam Raine’s suicide and the resulting lawsuit, suggesting that concern may now be even higher at the time of publication.

Many of the AI harms that are of most concern may be relevant due to their effect on relationships. When they are optimised for engagement, the manipulative intrusion of chatbots into the lives of psychologically vulnerable individuals, such as children, is damaging to relationships, especially within families.

Furthermore, this specific aspect of AI risk is germinating in fertile soil, as it were. Recent media attention has been clear in highlighting the negative effects of social media, especially for children. The public is primed to engage further on this issue. In our research, many respondents mentioned social media companies and their leaders as problematic. AI labs are themselves distrusted, with 42\% of the public believing that AI developers don’t have the public’s interests and safety in mind when developing models, against 33\% who believe they do. Technological change and AI are perceived by our respondents as something that happens \textit{to} them, a disempowering situation that heightens their sense of grievance and alienation (\citealt{57}).  

\section{Experimental validation: Message testing results}
To understand which topics truly drive action, we surveyed 1,063 people across the United States to identify which messages would make them most likely to take action, such as signing a petition or contacting their local representative.

This was done through testing 18 different messages, with three distinct messages split across six themes:

\begin{enumerate}
	\item Children
	\item Existential Risk
	\item Inequality
	\item Jobs
	\item Mental Health
	\item Weapons of Mass Destruction
\end{enumerate}

Using a polling technique called maximum difference scaling (MaxDiff), we measured the relative preference of these six different themes across the nation and different sub-groups, revealing which may most effectively motivate public engagement on AI.

In short, our testing found that messages grounded in tangible, lived risks, such as jobs and children, outperform abstract or existential concepts across all demographic cohorts. People are more likely to be mobilized most by what feels concrete and immediate in their lives.

\textbf{MaxDiff Methodology}

In a maximum difference scaling (MaxDiff) exercise, respondents are shown a subset of messages at a time and asked to choose the one they find most compelling and the one they find least compelling. Repeating this across different subsets reveals the relative strength of each message, helping to identify which resonates most strongly overall.

\begin{figure}[!ht]
    \centering
    \includegraphics[width=\textwidth]{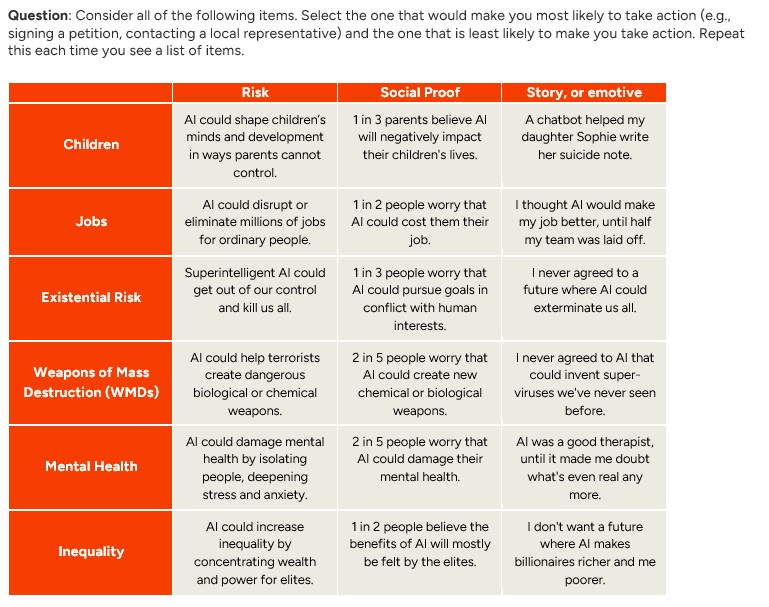}
    \caption{Research question and message grid for the MaxDiff message testing. Each respondent answered 15 questions, which each contained 4 of the messages chosen at random. They were asked the research question shown below.}
    \label{fig:2}
\end{figure}

Each respondent answered 15 questions, which each question displaying 4 of the 18 messages chosen at random. (See Fig. \ref{fig:2})

MaxDiff offers several advantages over more conventional message testing approaches, such as Likert ratings or simple rankings. Unlike rating scales, which often suffer from “top-box inflation” and make it difficult to distinguish between items, MaxDiff forces respondents to make clear trade-offs by selecting the most and least compelling option within a set. This approach reduces cognitive burden when compared to asking participants to rank a long list, while still producing robust interval-level data that show not only the order of preferences but also the relative strength of each message. 

Because the method is based on repeated choice exercises, it minimizes issues of scale bias (e.g., lenient versus harsh raters) and more closely mirrors real-world decision-making, where individuals must prioritize among competing options. As a result, MaxDiff provides a more reliable and discriminating measure of message effectiveness.

\begin{figure}[!ht]
    \centering
    \includegraphics[width=\textwidth]{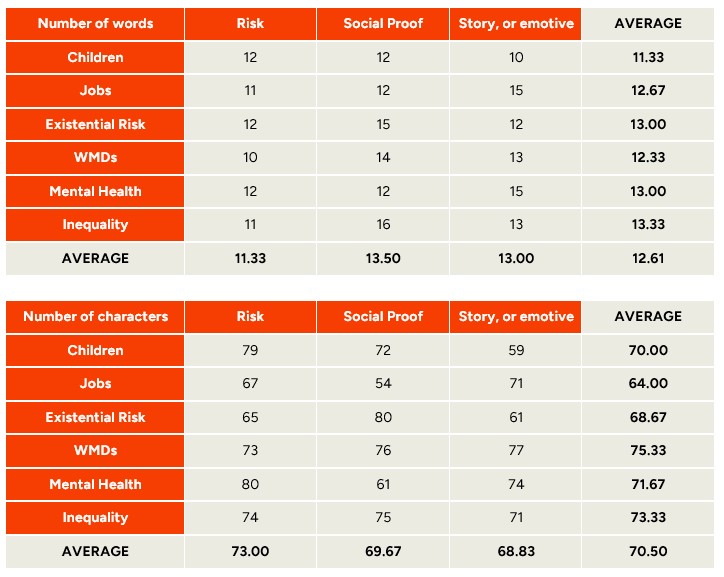}
    \caption{The tables below correspond to the equivalent message in the previous figure.}
    \label{fig:3}
\end{figure}

To help mitigate unconscious biases and other variables that may impact the MaxDiff scores, the messages were made as consistent as possible in both length and number of words, whilst ensuring that the key theme and tactic were honored. (See Fig. \ref{fig:3})

Data was collected via online surveys through a panel aggregator across five unique panel providers, with demographic quotas in place. There were 1063 respondents. Fieldwork took place on October 20th and 21st, 2025. After collection, the data was weighted by the following variables: gender, age, region, education, and 2024 Presidential vote.

\textbf{MaxDiff testing results}

\begin{figure}[!ht]
	\centering
	\includegraphics[width=\textwidth]{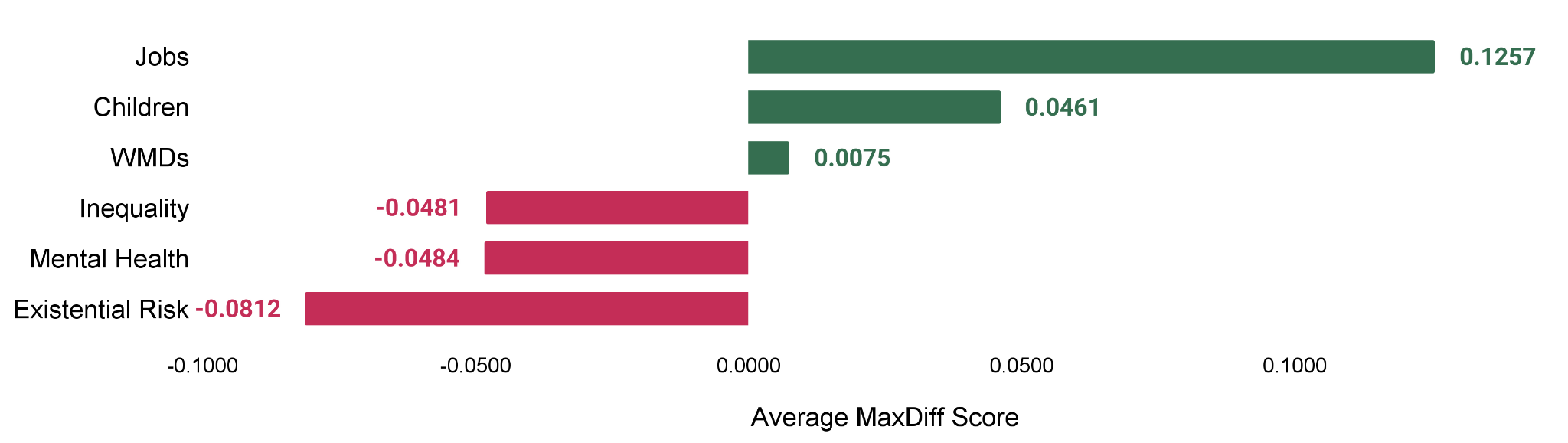}
	\caption{Message testing: average score by theme}
	\label{fig:4}
\end{figure}

When averaged by theme (see Fig. \ref{fig:4}), concerns about Jobs dominate, likely amplified by U.S. economic pressures such as inflation and rising living costs, which make employment disruption feel immediate and personal. 

Messages on the Children theme are also highly effective. Children’s safety is viewed as a bipartisan concern, giving it broader mobilizing power. This theme is intensely emotive, as seen in focus groups where participants repeatedly referenced tragedies such as Sewell Setzer and Adam Raine as examples of the worst outcomes. 

\begin{figure}[!ht]
	\centering
	\includegraphics[width=\textwidth]{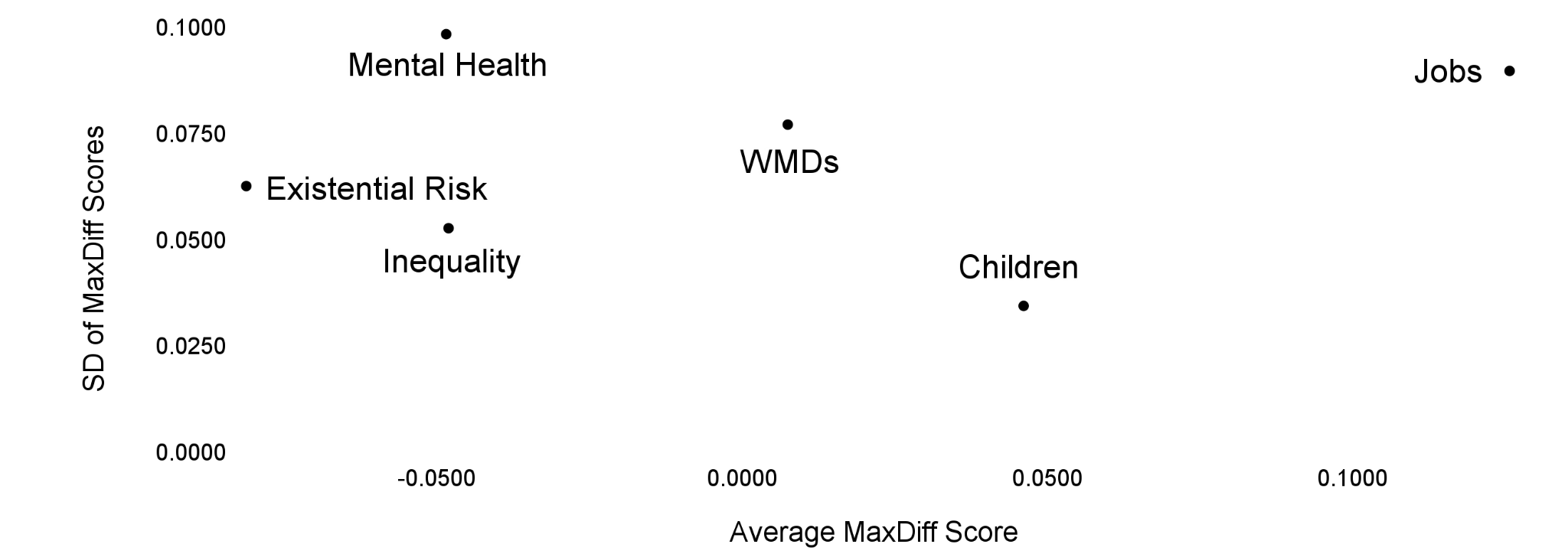}
	\caption{Average score and standard deviation by theme}
	\label{fig:5}
\end{figure}

Children’s messages also showed the lowest variance, indicating consistent effectiveness regardless of the message itself. (See Fig. \ref{fig:5})

In contrast, focus groups revealed that while people care deeply about jobs, they rarely connect personal employment risks to political action, often seeing it as their own responsibility to retrain or protect their position.

WMD-related messages exhibit modest engagement, though scores are likely heightened given current global security concerns. (Note that the message test ran in the days before the debut of “A House of Dynamite” on Netflix.)

Statements about Inequality, Mental Health, and Existential Risk were more likely to be picked as least persuasive. The Mental Health message average was significantly reduced, heavily influenced by one tested message with a storytelling approach: “AI was a good therapist, until it made me doubt what's even real anymore.” Excluding this statement, Mental Health scores average at zero, meaning the other two statements were selected as best and worst almost equally, reflected in the high standard deviation shown in the bottom chart. This suggests that future message exploration could perform well for some segments, but not others.

Existential Risk received the lowest scores overall, indicating that appeals focused on diffuse or long-term threats struggle to capture attention. Without immediate personal stakes, these abstract risks fail to mobilize concern or action from the general public.

\begin{figure}[!ht]
	\centering
	\includegraphics[width=\textwidth]{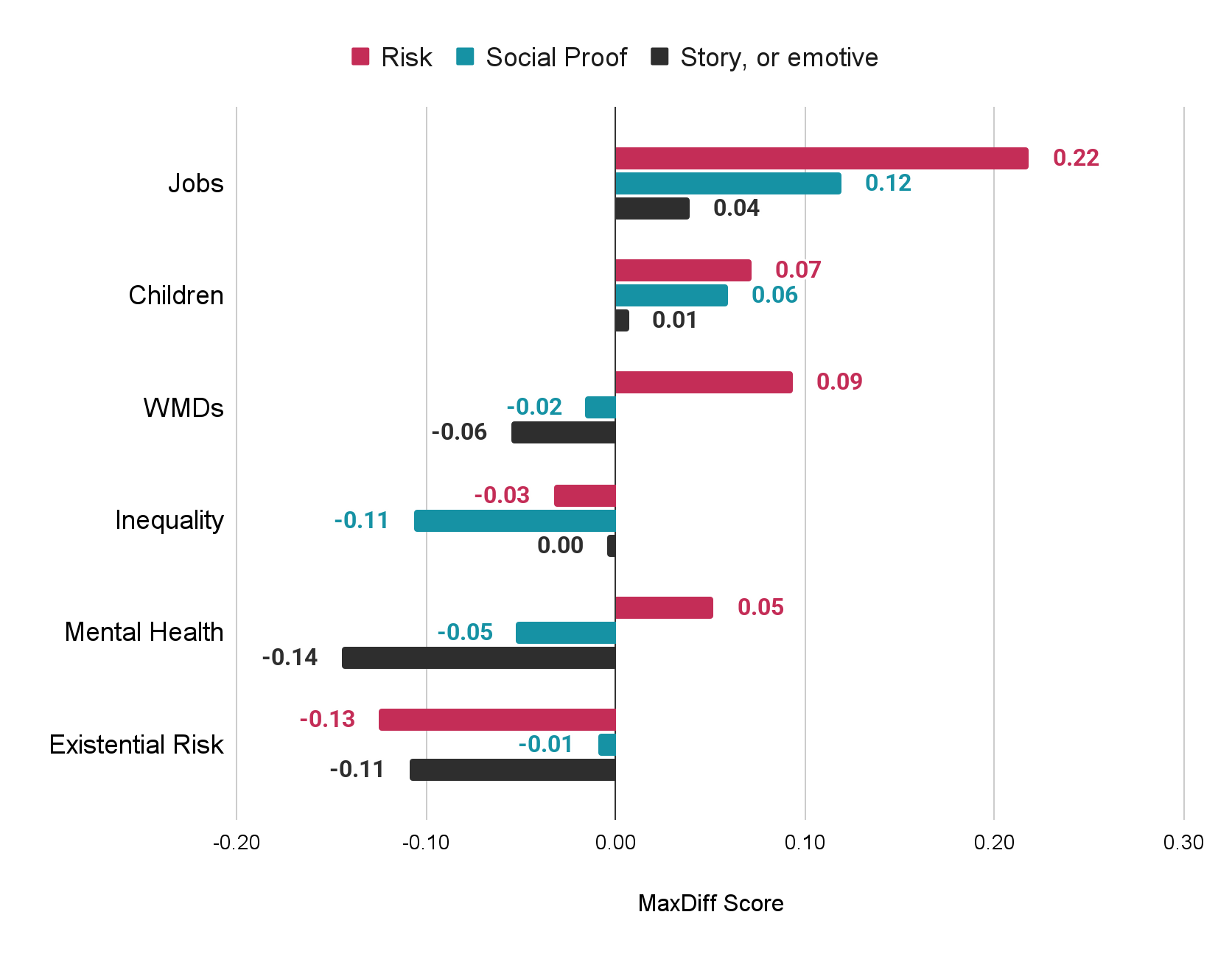}
	\caption{MaxDiff score by theme \& tactic}
	\label{fig:6}
\end{figure}

While messaging tactics matter, the underlying theme is ultimately more decisive in determining mobilization. Themes rooted in currently salient concerns, such as jobs and children, outperformed others, regardless of how they were framed. (See Fig. \ref{fig:6})

For both Jobs and Children statements, risk-framed tactics amplify resonance, further confirming that tangible messages are more effective than appeals to consensus, social proofing, or emotionally vivid but detached storytelling.

\begin{figure}[!ht]
	\centering
	\includegraphics[width=\textwidth]{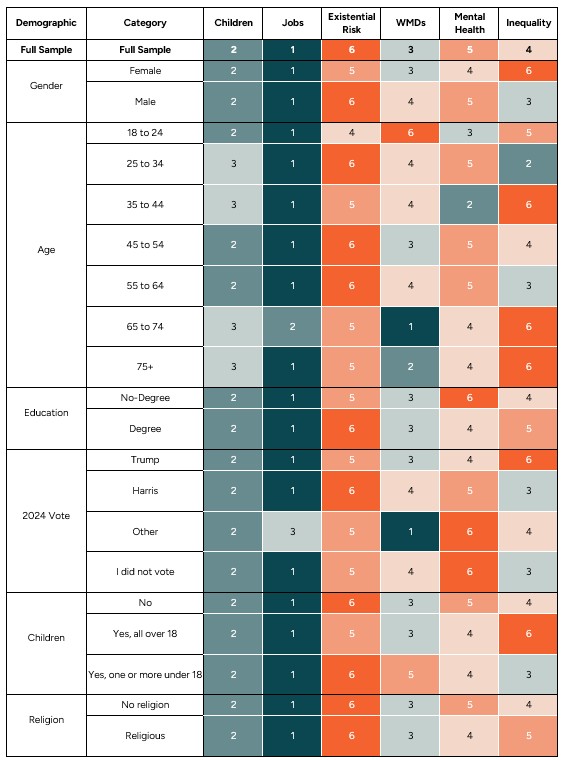}
	\caption{Rank of theme by demographic}
	\label{fig:7}
\end{figure}

Tangible, real-world consequences are more likely to work across all demographics. (See Fig. \ref{fig:7})

Older respondents, particularly those aged 65 and above with adult children, show the strongest engagement with Children and Jobs messages, suggesting that grandparent-age cohorts are especially concerned about AI’s social consequences. Child-related statements also perform well with non-parents, ranking as their second preference, and women over-index on these messages, highlighting a gender gap in care-oriented framings.

Younger respondents are more receptive to abstract or collective framings: those aged 25 to 34 respond most strongly to Inequality messages, while those aged 35 to 44 are more attentive to Mental Health statements. WMD messages resonate primarily with Trump voters and the oldest cohorts, reflecting a security-oriented worldview.

\textit{Existential Risk messaging is ineffective across all general public demographics}, ranking as the worst or second-worst theme for most groups and fourth out of six for 18–24-year-olds, underscoring the challenge of mobilizing attention around diffuse, long-term threats.

\textbf{Therefore, existential risk is not an effective theme for engaging the public. The two theme areas of Jobs \& Employment and Children \& Relationships offer opportunities to engage the public successfully on areas of AI policy.}  

\textit{Who do we reach out to that is easiest to convince?}

\section{The AI Publics}
\textit{Who should we be talking to about AI risk? Recent research shows which segments of the public are most likely to be receptive to messages about mitigating AI risk}

One of the key goals of Seismic’s research was to identify segments of the adult public who would be more responsive to messaging about AI risks and more likely to take civil society action compared to the general population, whilst identifying the themes to engage them on (\citealt{57}).

To this end, Seismic conducted a segmentation analysis across the respondents from all five polled countries. As part of the quantitative research, respondents were asked fifteen questions designed to be polarising. Each of these fifteen questions showed two diametrically opposed statements, asking respondents to position themselves on a scale between the two.

\begin{figure}[!ht]
	\centering
	\includegraphics[width=\textwidth]{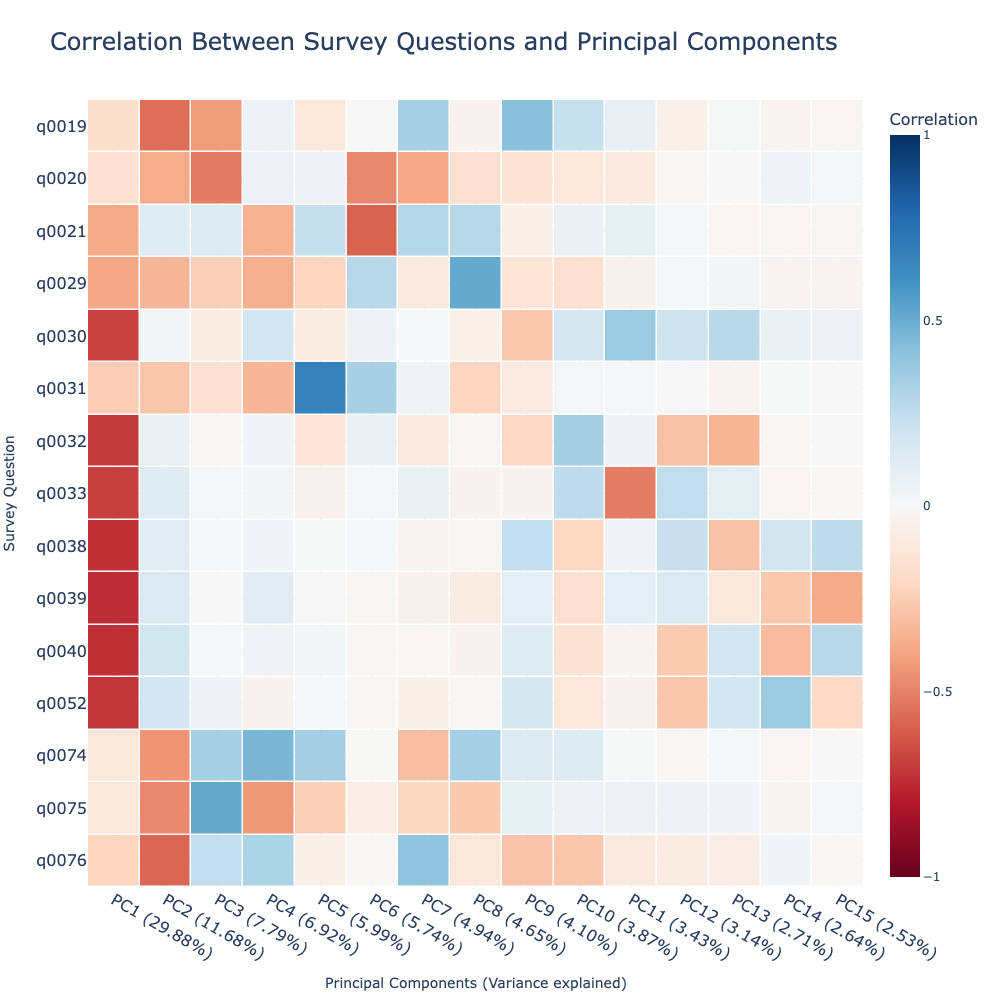}
	\caption{Correlation matrix of raw survey responses and PCA components (variance explained by each principal component shown in parentheses)}
	\label{fig:8}
\end{figure}

These fifteen questions were input into a principal component analysis (PCA), after being scaled (standard deviation of each slider column forced to be equal to 1) and centered (mean forced to be 0), and weighted by the sample weights so that the correlation matrix is representative of true population proportions for each country. For this process, it was decided to give each of the polled countries equal importance. Each principal component, i.e., custom indices, is orthogonal or uncorrelated with the other. Each principal component is a linear combination of the original variables, aligned with an eigenvector of the empirical weighted covariance matrix. The corresponding eigenvalue gives the variance explained by that component. By ordering eigenvalues from largest to smallest, PCA frontloads the variance into the first components, so the first few capture most of the information in the data. Public attitudes towards AI and the world at large are a multi-faceted issue; the PCA process is an effective way to combine multiple questions whilst also accounting for columns with minimal variance or those that correlate strongly with each other (see Fig. \ref{fig:8}).

\begin{align*}
	&\text{1. Data matrix:} \quad 
	X \in \mathbb{R}^{n \times p}, \quad X = \{ x_{ij} \} \\[6pt]
	&\text{2. Standardization (mean = 0, sd = 1 per column):} \quad
	Z_{ij} = \frac{x_{ij} - \mu_j}{\sigma_j}, \quad 
	\mu_j = \frac{1}{n} \sum_{i=1}^n x_{ij} \\[6pt]
	&\text{3. Apply survey weights:} \quad
	\tilde{Z}_{i \cdot} = \sqrt{w_i} \, Z_{i \cdot} \\[6pt]
    &\text{Matrix form: } \tilde{Z} = W^{1/2} Z, \quad W = \mathrm{diag}(w_1, \dots, w_n) \\[6pt]
	&\text{4. Weighted covariance matrix:} \quad
	\Sigma_w = \tilde{Z}^\top \tilde{Z} \\[6pt]
	&\text{5. Eigen-decomposition (PCA):} \quad
	\Sigma_w v_k = \lambda_k v_k, \quad k = 1, \dots, p \\[6pt]
	&\text{6. Variance explained ratio:} \quad
	\text{VarExplained}(k) = \frac{\lambda_k}{\sum_{j=1}^p \lambda_j} \\[6pt]
	&\text{7. Principal component scores:} \quad
	\text{PC}_{k} = \tilde{Z} v_k
\end{align*}

There are benefits to processing the data in this manner before the segmentation:

\begin{enumerate}
    \item Enhanced interpretability through indices: Public attitudes toward AI are multifaceted, hence the need to ask fifteen slider questions in the survey. However, these attitudes are often correlated. PCA combines related questions into principal components, creating latent indices that capture coherent dimensions of opinion. For example, one component might aggregate questions about AI and the economy into a single axis, simplifying the interpretation of clusters.
    \item Mitigation of bias from correlated variables: Highly correlated survey items can disproportionately influence clustering if left unprocessed. For instance, if the survey included five times as many questions about civic engagement as about AI’s economic impact, clusters would be unduly driven by the civic questions. The intention is to capture a multifaceted view driven by empirical variance in the public, rather than the researcher’s bias on which questions to include.
    \item Dimensionality reduction and improved cluster geometry: PCA reduces the number of input variables while preserving the most informative variation, focusing the Gaussian mixture model (GMM) on the key structure in the data. At the same time, it transforms the feature space so that the axes are uncorrelated, producing cluster shapes that better match the Gaussian assumptions of the model and improving the stability and accuracy of the Expectation Maximization (EM) steps.
\end{enumerate}

As stated above, the segmentation algorithm used after the PCA was a Gaussian mixture model (GMM). GMM is a probabilistic clustering algorithm that assumes segments of the survey respondents and that they can be modelled as a Gaussian distribution in the transformed feature space. The GMM is fitted using an implementation of the Expectation Maximization algorithm (EM) to maximise the log-likelihood of the data under the mixture model, i.e., the likelihood that the found distributions generate the empirically observed survey data. The process repeats until convergence. This algorithm has some major benefits when compared against other segmentation algorithms used in modeling public opinion, particularly k-means:

\begin{enumerate}
    \item GMMs allow found segments to vary in size, shape, and orientation. This can help capture niche groups that are often amalgamated in k-means
    \item 	k-means uses a distance-based loss function that has no inherent understanding that principal components have different levels of importance, based on the proportion of the original variance that they explain
\end{enumerate}
 	
\begin{align*}
	&\text{1. Mixture model:} \quad
	p(x) = \sum_{k=1}^K \pi_k \, \mathcal{N}(x \mid \mu_k, \Sigma_k) \\[6pt]
	&\text{2. Gaussian component density:} \quad
	\mathcal{N}(x \mid \mu_k, \Sigma_k) 
	= \frac{1}{(2\pi)^{d/2} |\Sigma_k|^{1/2}} 
	\exp\!\left( -\tfrac{1}{2} (x - \mu_k)^\top \Sigma_k^{-1} (x - \mu_k) \right) \\[6pt]
	&\text{3. Mixing coefficients (priors):} \quad
	\pi_k \geq 0, \quad \sum_{k=1}^K \pi_k = 1 \\[6pt]
	&\text{4. Log-likelihood (objective function):} \quad
	\ell(\pi, \mu, \Sigma) 
	= \sum_{i=1}^n \log \left( \sum_{k=1}^K \pi_k \, \mathcal{N}(x_i \mid \mu_k, \Sigma_k) \right) \\[6pt]
	&\text{5. Posterior probabilities (responsibilities, E-step):} \quad
	\gamma_{ik} = 
	\frac{\pi_k \, \mathcal{N}(x_i \mid \mu_k, \Sigma_k)}{\sum_{j=1}^K \pi_j \, \mathcal{N}(x_i \mid \mu_j, \Sigma_j)} \\[6pt]
	&\text{6. M-step parameter updates:} \quad
	\pi_k^{\text{new}} = \frac{1}{n} \sum_{i=1}^n \gamma_{ik}, 
	\quad
	\mu_k^{\text{new}} = \frac{\sum_{i=1}^n \gamma_{ik} x_i}{\sum_{i=1}^n \gamma_{ik}} \\[6pt]
	&\qquad\qquad
	\Sigma_k^{\text{new}} = \frac{\sum_{i=1}^n \gamma_{ik} (x_i - \mu_k)(x_i - \mu_k)^\top}{\sum_{i=1}^n \gamma_{ik}}
\end{align*}

The outputs of the GMM are the posterior probabilities that any survey respondent belongs to any mixture component, i.e., each Gaussian distribution in the model. The analysis was run with twenty mixture components. For interpretation and including segments in the survey cross-tabulations, each respondent was assigned to the segment corresponding to the component with the highest posterior probability of membership. Twenty segments were chosen based on the Bayesian Information Criterion (BIC), which balances the model’s fit, maximized log-likelihood against its complexity (the number of parameters), in order to prevent overfitting.

\begin{equation*}
    \text{BIC} = -2 \log \hat{L} + k \log n
\end{equation*}

\begin{figure}[!ht]
	\centering
	\includegraphics[width=\textwidth]{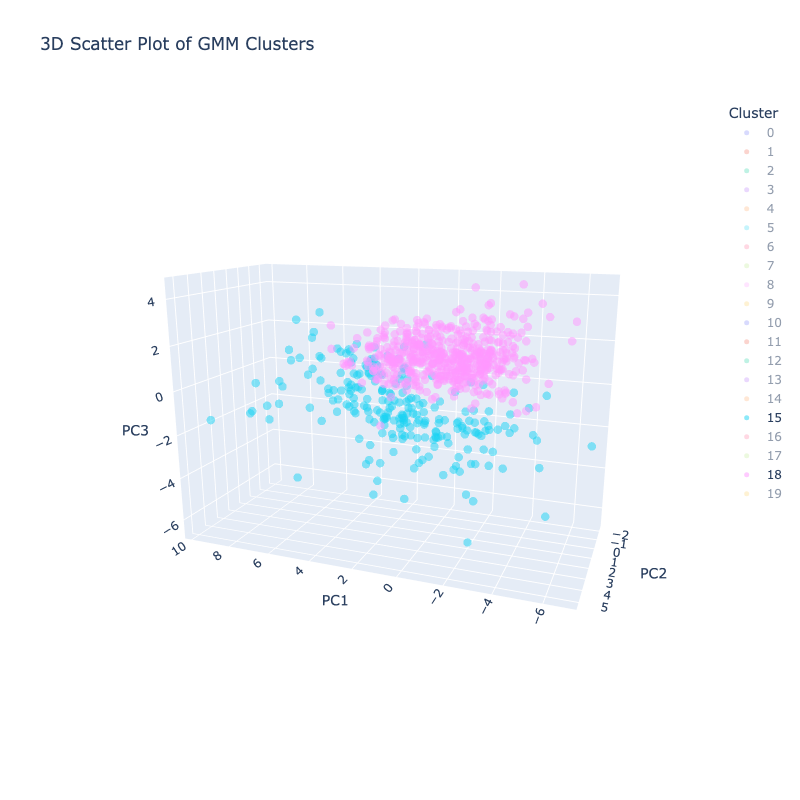}
	\caption{Survey respondents for selected clusters, displayed on the first three principal components}
	\label{fig:9}
\end{figure}

A visual example of select segments is provided in the Figures section (see Fig. \ref{fig:9}). Each point on the scatterplot represents a survey respondent, coloured based on their segment and positioned based on their values for the first three principal components. Note that the first three components represent 49\% of the total variance.

Five of these twenty segments were isolated due to their higher concern about particular aspects of AI, and/or their perceived job exposure to AI, coupled with a higher propensity to take civic action. These five are covered in the full report (\citealt{57}). The following concentrates on two of these five: “Tech-Positive Urbanites” and “World Guardians” due to, but not limited to, their high turnout in elections, propensity to take civic action, awareness about AI, perceived risk that AI will have on their employment circumstances, and attitudes towards AI’s impact on social matters.

\textbf{Tech-Positive Urbanites }

The Tech-Positive Urbanites represent 20.2 million people extrapolated across all participating nations. Tech-Positive Urbanites are more likely to be highly educated, politically engaged, actively following news and current events (favoring international sources for more balanced reporting), while participating in protests and civic actions much more often than the average citizen of their country. 

They are, as the name suggests, avid users of AI technology, more likely than the average person to trust AI to, for example, manage their finances or mentor their children. However, they are set apart by also being acutely worried about AI risks, especially through the lens of their own jobs – much more so than the average citizen. Most have mid-level white-collar jobs, often in technical roles.

Many of their deepest concerns about AI revolve around its impact on their jobs, financial stability, and future well-being. They are twice as likely as the average respondent to consider their current job role as being exposed to AI disruption, choosing “Very high impact: Almost everything I do could soon be done by AI.” 

In qualitative research, through focus groups, respondents from this segment deepened this theme, exhibiting a very high level of concern about AI’s integration into the economy. Critically, they placed the challenge of dealing with AI within a wider narrative of loss of control (\citealt{57}). Some indicated a very high level of awareness of how AI might affect them unequally:

\begin{itemize}
    \item	“A capitalist system cannot run without the exploitation of free or cheap labor. Somebody has to be the have-nots for there to be the haves” (respondent from Florida) 
    \item	“That's one of the biggest problems with AI today. A lot of the rich people, or the tech bros, they… they just see dollar signs, dollar signs, money, money, money, money, money. And… they're like… They just wanna…make everything AI.” (respondent from Illinois)
\end{itemize}

Other areas of concern include AI’s effect on bias and misinformation, its impact on children and relationships, and, despite recognizing AI’s benefits, they were concerned about maintaining the importance of human intervention, empathy, and critical thinking.

\textbf{World Guardians}

The second high-priority segment, representing 31.2 million people, has a somewhat broader focus of concern than the previous group. While the Tech-Positive Urbanites are focused on themselves and their families, the World Guardians are much more likely to express a concern for issues they see as important in the wider world. They are generally more affluent, skew slightly older, and lean socially progressive. They are highly engaged with news and current events, and while they are socially progressive, they are somewhat technologically conservative, favoring traditional national and local TV news sources. They show strong civic participation through high voter turnout and signing petitions. 

Among their distinguishing attitudes, World Guardians are almost twice as likely to worry that future generations will have it harder with AI as a risk to society as a whole, despite feeling that their own jobs are not as vulnerable to AI disruption. 

For this group, AI exists within a complex of concerns; engagement with prominent social issues is an important social signifier for this group. They worry about climate change more than the average respondent, and also consider matters such as the existential risks of AI when it comes to its addition to pre-established existential risks to society, such as AI decision-making in warfare (57\%) and in politics (54\%). Half are worried about AI replacing human relationships (49\%), and they see AI as likely to worsen existing societal problems, particularly climate change. They tend to mistrust AI developers and are the most likely segment (60\%) to think there is currently not enough regulation around AI.

Furthermore, they expressed numerous concerns about how AI affects various aspects of our everyday life, such as job loss and security, deepfakes, AI use in scams, and the spread of misinformation, as well as concerns about its impact on physical and mental health, and its potential to replace human connection.

More information on these groups is available within the full Seismic report, On the Razor’s Edge: AI vs. Everything We Care About.

\textbf{These audience segments are the easiest to mobilize to take civil society action on AI harms and put pressure on policymakers.}

\textit{But will this lead to the policies we need to mitigate the most grievous AI harms?}

\section{A pathway to Policy Effectiveness}  
\textit{If the publics are mobilized around policies addressing immediate harms from AI, can this momentum extend to governance frameworks that also mitigate catastrophic risks? The answer is not automatic, but there is a plausible pathway.}

Our research and that of others show that the public does not view AI as a single issue with a single solution, but instead understands it through multiple interwoven concerns, including relationships, employment, mental health, misinformation, and more (\citealt{57}). This is consistent with prior research on “issue bundling” in public opinion, where heterogeneous but related concerns accumulate into broader policy demands (\citealt{20}).

This dynamic is well illustrated with the example of environmental policy. Highly disparate issues such as air quality, biodiversity, and energy security eventually coalesced into a durable policy agenda under the banner of “the environment.” 

This process cannot be assumed to be automatic. The analogy, therefore, functions as a heuristic, not a prediction. The challenge is to ensure that action on proximate concerns translates into structural reforms rather than superficial or populist fixes.

Public pressure across a wide range of concerns can create what we might call “policy demand.” Policymakers, facing a chorus of different constituencies, become incentivized to act—and the policies they enact can extend beyond the narrow issue that first provoked public outrage.

For example, let us consider the current debate on AI-driven suicides, which has given rise to concentrated interest from the policy community (see, e.g., \citealt{44}). This is a tragic and specific issue. Yet the policies needed to mitigate it can lead to broad and structural policy change:

\begin{itemize}
    \item Mandatory pre-deployment safety testing
    \item Independent audits and verification
    \item Liability and accountability frameworks
    \item Transparency requirements and whistleblower protections
\end{itemize}
 
These measures, though they might plausibly be introduced to address immediate harms, are generalizable. They also build institutional capacity that can apply to less visible or less tangible risks, including catastrophic ones. 

In other words, the same regulatory scaffolding designed to prevent narrow harms can also be built on to mitigate systemic threats.

That said, this pathway is contingent. Mobilization around proximate harms could stall at shallow interventions, such as age restrictions on chatbots or symbolic bans, that fail to build broader governance capacity. Whether the “spillover” effect materializes depends on deliberate action by communicators and policymakers. This means not resting with the achievement of initial policy successes, but building on them, framing each proximate issue not as an isolated grievance but as evidence of the need for systemic AI governance.

\textit{In short, mobilizing publics around everyday harms is not sufficient in itself, but it is necessary. Without salience, policymakers will not act against entrenched interests; with it, they may enact measures that, if designed structurally, build resilience against the full spectrum of AI risks, including catastrophic ones.}

\section{Limitations and Future Research}
Several limitations constrain this analysis. 

First, our application of psychological theories such as Terror Management Theory and the Self-Reference Effect to AI risk communication is inferential. While consistent with prior research in other domains, we lack direct experimental evidence demonstrating their operation in the context of AI specifically. Future work should test these mechanisms through controlled studies that expose participants to alternative framings of AI risk and measure attitudinal and behavioral outcomes.

Second, while our survey data cover five countries, they represent a narrow slice of the global population, weighted toward higher-income democracies. The majority of similar survey data is even narrower in scope (i.e., US only), but public attitudes toward AI may diverge substantially in regions with different political cultures, media ecosystems, and lived experiences of technology.  Comparative research across the Global South is necessary to validate or refine our segmentation framework.

From a segmentation point of view, each of the five countries was given the same importance, despite the countries having a non-uniform impact on the pace of AI innovation or the setting of regulations. Furthermore, not all countries have the same population size.

Third, the use of Google Trends data as a proxy for public salience is illustrative but limited. Search patterns are volatile, subject to media shocks, and not necessarily predictive of sustained mobilization. Longitudinal surveys and behavioral measures of civic engagement, such as donation or membership data from civil society organizations, would provide a stronger foundation for tracking shifts in salience over time. 

Fourth, the message testing we conducted was necessarily limited in the number of messages that could be compared, when considered against the vast possibility space of variation in messages, themes, and messaging styles. There is a clear opportunity for further testing here. This could also incorporate other elements, e.g., imagery. There is ample opportunity to extend and expand the scope of message testing. A test-and-learn approach will be necessary to extend the findings of this paper, and to optimize individual pieces of content or campaigns launched in the future. 

Finally, our theory of change, which posits that mobilization around proximate harms can generate structural governance with spillover effects into catastrophic risk mitigation, remains contingent. It depends on communicators and policymakers deliberately linking everyday harms to systemic reforms. Under some political conditions, campaigns focused on immediate harms may stall at shallow or symbolic interventions (e.g., parental controls for chatbots, content warnings) that pacify public concern without building institutional capacity. Worse, they could trigger counter-mobilization by entrenched interests, reframing AI risk as moral panic or regulatory overreach. Future work should therefore focus on maintaining and increasing the pace and ambition of regulatory change in the aftermath of initial successes on proximate harms. Mechanisms likely include the roles of intermediaries (such as media framing, NGO advocacy, and political entrepreneurs) and the degree of institutional openness to reform.

By addressing these gaps, subsequent studies can better establish whether and how proximate concerns can serve as a reliable pathway to robust governance of catastrophic AI risks.

\section{Conclusion}
Public support is a necessary condition for durable AI governance. In democracies, policy outcomes are shaped not only by elite preferences and lobbying but also by the salience of issues in the public mind (\citealt{15}; \citealt{72}). Currently, AI ranks low on public priority lists, despite widespread concern when prompted. This creates a gap between latent worry and political viability.

Attempts to build mobilization around existential or catastrophic risks of AI have, to date, been unsuccessful. As shown in Section 2, this is consistent with established psychological dynamics: mortality avoidance, poor intuitive grasp of exponential growth, and the lack of personal reference points. It is also consistent with public opinion trends, which indicate stagnating concern about “loss of control” scenarios, and persistently low public searches or engagement around extinction risk.

In this context, attempts to build a popular movement specifically around the catastrophic risks of AI are likely to be difficult, or even counterproductive. In this sense, we agree with recent calls to avoid building a public movement narrowly focused on AI safety (e.g., \citealt{40}). 

However, this is not the only way to build a movement pressing for policy change. Our analysis suggests that this does not mean AI risk communication \textit{in the broad sense} is inherently counterproductive. In fact, we have shown that some kind of public movement is necessary to bring about policy change focused on mitigating AI risks in a robust fashion. However, for this to be successful, a necessary condition is that engagement must begin with harms that people already perceive as relevant, such as threats to jobs, dignity, mental health, and relationships. These proximate concerns can generate the salience required to move AI up the policy agenda. Only then are policymakers likely to take the required actions in the face of countervailing pressure from elite and institutional factors.

The critical question is whether this type of public mobilization on proximate harms can also generate governance structures capable of addressing broader, catastrophic risks. We have argued that it can, but only if campaigns and communicators deliberately link proximate harms to systemic regulatory solutions. Measures such as pre-deployment testing, independent audits, liability frameworks, and transparency rules may be prompted by narrow crises—such as chatbot-induced suicides—but once institutionalized, they can extend to mitigate broader risks. 

This “spillover” dynamic is not guaranteed, but historical precedents suggest it is possible under the right conditions and with concerted effort. In contrast, the evidence suggests that this is not true of focus on existential or catastrophic risks.

Therefore, the task for communicators and policymakers is twofold:

\begin{enumerate}
    \item \textbf{Engage the publics} on proximate harms in ways that are emotionally resonant and relatable;
    \item \textbf{Translate that engagement} into structural governance reforms that are flexible and scalable enough to address the full spectrum of AI risks, including existential ones.
\end{enumerate}

AI’s trajectory is still unfolding, but its integration into society is accelerating. Ensuring that governance keeps pace will require deliberate, strategic communication that connects everyday concerns to systemic policy needs. 

Done effectively, this approach provides the best available pathway to mobilizing publics, counterbalancing entrenched interests, and creating durable institutions capable of safeguarding against both present harms and future catastrophic risks.

\section*{Acknowledgments }
The authors wish to thank Samuel Härgestam and Sören Mindermann for their valuable input and questions. Any errors or omissions are the sole responsibility of the principal author. 

\bibliographystyle{unsrtnat}
\bibliography{references}  

@article{1,
  author       = {Aaronson, Scott},
  title        = {In search of AI psychosis},
  journal      = {Astral Codex Ten},
  year         = {2025},
  month        = {August},
  note         = {Retrieved September 9, 2025 from \url{https://www.astralcodexten.com/p/in-search-of-ai-psychosis}}
}

@misc{2,
  author       = {Altman, Sam},
  title        = {The gentle singularity},
  year         = {2025},
  month        = {June},
  howpublished = {Blog Post},
  note         = {Retrieved September 23, 2025 from \url{https://blog.samaltman.com/the-gentle-singularity}}
}

@misc{3,
  author       = {Amodei, Dario},
  title        = {Machines of loving grace},
  year         = {2024},
  month        = {October},
  howpublished = {Blog Post},
  note         = {Retrieved September 23, 2025 from \url{https://www.darioamodei.com/essay/machines-of-loving-grace}}
}

@article{4,
  author       = {An, E. and Wennberg, E. and Nissim, R. and Lo, C. and Hales, S. and Rodin, G.},
  title        = {Death talk and relief of death-related distress in patients with advanced cancer},
  journal      = {BMJ Supportive \& Palliative Care},
  year         = {2020},
  volume       = {10},
  pages        = {e19},
  doi          = {10.1136/bmjspcare-2016-001277},
  url          = {https://doi.org/10.1136/bmjspcare-2016-001277}
}

@book{5,
  author       = {Asimov, Isaac},
  title        = {Robot visions},
  publisher    = {Penguin Books},
  year         = {1990}
}

@article{6,
  author       = {Barabas, Jason},
  title        = {Democracy’s denominator: Reassessing responsiveness with public opinion on the national policy agenda},
  journal      = {Public Opinion Quarterly},
  year         = {2016},
  volume       = {80},
  number       = {2},
  pages        = {437--459},
  doi          = {10.1093/poq/nfv082},
  url          = {https://doi.org/10.1093/poq/nfv082}
}

@article{7,
  author       = {Beattie, Geoffrey and Sale, Laura and McGuire, Louise},
  title        = {An inconvenient truth? Can a film really affect psychological mood and our explicit attitudes towards climate change?},
  journal      = {Semiotica},
  year         = {2011},
  volume       = {187},
  number       = {1-4},
  pages        = {105--125},
  doi          = {10.1515/semi.2011.066},
  url          = {http://dx.doi.org/10.1515/semi.2011.066}
}

@article{8,
  author       = {Bengio, Yoshua and Mindermann, Sören and Privitera, Davide and Besiroglu, Tamay and Bommasani, Rishi and Casper, Stephen and Choi, Yejin and Fox, Peter and Garfinkel, Ben and Goldfarb, Daniel and Heidari, Hoda and Ho, Andrew and Kapoor, Sayash and Khalatbari, Leila and Longpre, Shayne and Manning, Sam and Mavroudis, Vasileios and Mazeika, Mantas and Michael, Julian and Zeng, Yuling},
  title        = {International AI safety report},
  journal      = {arXiv},
  year         = {2025},
  note         = {\url{https://doi.org/10.48550/arXiv.2501.17805}},
  doi          = {10.48550/arXiv.2501.17805}
}

@misc{9,
  author       = {Bengio, Yoshua},
  title        = {The catastrophic risks of AI — and a safer path},
  year         = {2025},
  month        = {May},
  howpublished = {Video},
  publisher    = {TED},
  note         = {Retrieved September 12, 2025 from \url{https://youtu.be/qe9QSCF-d88?si=VVHVvjeLq1D6LbCa}}
}

@article{10,
  author       = {Bengio, Yoshua and Hinton, Geoffrey and Yao, Andrew and Song, Dawn and Abbeel, Pieter and Darrell, Trevor and Harari, Yuval Noah and Zhang, Yu-Qing and Xue, Lei and Shalev-Shwartz, Shai and Hadfield, Gillian and Clune, Jeff and Maharaj, Tanya and Hutter, Frank and Baydin, Atilim Gunes and McIlraith, Sheila and Gao, Qian and Acharya, Abhishek and Krueger, David and Mindermann, Sören},
  title        = {Managing extreme AI risks amid rapid progress},
  journal      = {Science},
  year         = {2024},
  volume       = {384},
  pages        = {842--845},
  doi          = {10.1126/science.adn0117},
  url          = {https://doi.org/10.1126/science.adn0117}
}

@article{11,
  author       = {Boissat, Laurie and Thomas-Walters, Laura and Veríssimo, Diogo},
  title        = {Nature documentaries as catalysts for change: Mapping out the ‘Blackfish Effect’},
  journal      = {People and Nature},
  year         = {2021},
  volume       = {3},
  number       = {9},
  pages        = {1179--1192},
  doi          = {10.1002/pan3.10221},
  url          = {https://doi.org/10.1002/pan3.10221}
}

@article{12,
  author       = {Bromley-Trujillo, Rebekah and Poe, Jeremy},
  title        = {The importance of salience: Public opinion and state policy action on climate change},
  journal      = {Journal of Public Policy},
  year         = {2020},
  volume       = {40},
  number       = {2},
  pages        = {280--304},
  doi          = {10.1017/S0143814X18000375},
  url          = {https://doi.org/10.1017/S0143814X18000375}
}

@misc{13,
  author       = {Brynjolfsson, Erik and Chandar, Bharat and Chen, Ruidi},
  title        = {Canaries in the coal mine? Six facts about the recent employment effects of artificial intelligence},
  year         = {2025},
  howpublished = {Working paper},
  note         = {Latest version available at \url{https://digitaleconomy.stanford.edu/publications/canaries-in-the-coal-mine}. Retrieved September 23, 2025.}
}

@article{14,
  author       = {Bullock, John B. and Pauketat, Julius V. T. and Huang, Hong and Wang, Yu Fang and Anthis, Jacy Reese},
  title        = {Public opinion and the rise of digital minds: Perceived risk, trust, and regulation support},
  journal      = {Public Performance \& Management Review},
  year         = {2025},
  pages        = {1--32},
  doi          = {10.1080/15309576.2025.2495094},
  url          = {https://doi.org/10.1080/15309576.2025.2495094}
}

@article{15,
  author       = {Burstein, Paul},
  title        = {The impact of public opinion on public policy: A review and an agenda},
  journal      = {Political Research Quarterly},
  year         = {2003},
  volume       = {56},
  number       = {1},
  pages        = {29--40},
  doi          = {10.2307/3219881},
  url          = {https://doi.org/10.2307/3219881}
}

@incollection{16,
  author       = {Capstick, Edward},
  title        = {Chapter 8: Public opinion},
  booktitle    = {The HAI Artificial Intelligence Index Report},
  publisher    = {Stanford University Human-Centered AI Institute},
  year         = {2025},
  url          = {https://hai.stanford.edu/assets/files/hai_ai-index-report-2025_chapter8_final.pdf}
}

@misc{17,
  author       = {{Center for AI Safety}},
  title        = {Statement on AI Risk},
  year         = {2023},
  note         = {Retrieved September 11, 2025 from \url{https://aistatement.com/}}
}

@misc{18,
  author       = {{Centre for Future Generations}},
  title        = {Preparing for AI labour shocks should be a resilience priority for Europe},
  year         = {2025},
  url          = {https://cfg.eu/ai-labour-shocks/}
}

@article{19,
  author       = {Choi, Sujin},
  title        = {The two-step flow of communication in Twitter-based public forums},
  journal      = {Social Science Computer Review},
  year         = {2015},
  volume       = {33},
  number       = {6},
  pages        = {696--711},
  doi          = {10.1177/0894439314556599},
  url          = {https://doi.org/10.1177/0894439314556599}
}

@article{20,
  author       = {Cuppen, Eefje and Ejderyan, Olivier and Pesch, Udo and Spruit, Shannon and van de Grift, Eliza and Correljé, Aad and Taebi, Behnam},
  title        = {When controversies cascade: Analysing the dynamics of public engagement and conflict in the Netherlands and Switzerland through “controversy spillover”},
  journal      = {Energy Research \& Social Science},
  year         = {2020},
  volume       = {68},
  pages        = {Article 101593},
  doi          = {10.1016/j.erss.2020.101593},
  url          = {https://doi.org/10.1016/j.erss.2020.101593}
}

@book{21,
  author       = {Dinello, Daniel},
  title        = {Technophobia!: Science fiction visions of posthuman technology},
  publisher    = {University of Texas Press},
  year         = {2005},
  doi          = {10.7560/709546},
  url          = {https://doi.org/10.7560/709546}
}

@article{22,
  author       = {Dietrich, Daniel},
  title        = {For America to rise it’s a matter of Black lives / And we gonna free them, so we can free us: 13th and social justice documentaries in the age of “fake news.”},
  journal      = {Pacific Coast Philology},
  year         = {2019},
  volume       = {54},
  number       = {2},
  pages        = {220--251},
  doi          = {10.5325/pacicoasphil.54.2.0220},
  url          = {https://doi.org/10.5325/pacicoasphil.54.2.0220}
}

@misc{23,
  author       = {{Department for Science, Innovation and Technology}},
  title        = {International survey of public opinion on AI safety},
  year         = {2023},
  note         = {Retrieved September 9, 2025 from \url{https://www.gov.uk/government/publications/international-survey-of-public-opinion-on-ai-safety}}
}

@misc{24,
  author       = {{Edelman}},
  title        = {Trust barometer 2025},
  year         = {2025},
  note         = {Retrieved September 17, 2025 from \url{https://www.edelman.com/trust/2025/trust-barometer}}
}

@misc{25,
  author       = {{European Commission}},
  title        = {Commission survey shows most Europeans support use of artificial intelligence in the workplace},
  year         = {2025},
  month        = {February 13},
  note         = {Retrieved September 29, 2025 from \url{https://employment-social- affairs.ec.europa.eu/news/commission-survey-shows-most-europeans-support-use-artificial-intelligence-workplace}}
}

@misc{26,
  author       = {{European Commission}},
  title        = {Strategic foresight report 2025},
  year         = {2025},
  month        = {September 9},
  note         = {Retrieved September 29, 2025 from \url{https://commission.europa.eu/strategy-and-policy/strategic-foresight/2025-strategic-foresight-report_en}}
}

@misc{27,
  author       = {Fenwick, Chris and Qureshi, Zuber},
  title        = {Risks from power-seeking AI systems},
  publisher    = {80{,}000 Hours},
  year         = {2022},
  note         = {Retrieved September 12, 2025 from \url{https://80000hours.org/problem-profiles/risks-from-power-seeking-ai/}}
}

@book{28,
  author       = {Friedman, Ted},
  title        = {Electric dreams: Computers in American culture},
  publisher    = {New York University Press},
  year         = {2005}
}

@article{29,
  author       = {Giuntella, Osea and König, Johannes and Stella, Luca},
  title        = {Artificial intelligence and the well-being of workers},
  journal      = {Scientific Reports},
  year         = {2025},
  volume       = {15},
  pages        = {20087},
  url          = {https://www.nature.com/articles/s41598-025-98241-3}
}

@incollection{30,
  author       = {Greenberg, Jeff and Pyszczynski, Tom and Solomon, Sheldon},
  title        = {The causes and consequences of a need for self-esteem: A terror management theory},
  booktitle    = {Public self and private self},
  editor       = {Baumeister, Roy F.},
  pages        = {189--212},
  publisher    = {Springer-Verlag},
  year         = {1986},
  doi          = {10.1007/978-1-4613-9564-5_10},
  url          = {https://doi.org/10.1007/978-1-4613-9564-5_10}
}

@incollection{31,
  author       = {Greenberg, Jeff and Solomon, Sheldon and Pyszczynski, Tom},
  title        = {Terror management theory of self-esteem and cultural worldviews: Empirical assessments and conceptual refinements},
  booktitle    = {Advances in experimental social psychology},
  editor       = {Zanna, Mark P.},
  volume       = {29},
  pages        = {61--139},
  publisher    = {Academic Press},
  year         = {1997},
  doi          = {10.1016/S0065-2601(08)60016-7},
  url          = {https://doi.org/10.1016/S0065-2601(08)60016-7}
}

@article{32,
  author       = {Grossmann, Matt},
  title        = {Interest group influence on U.S. policy change: An assessment based on policy history},
  journal      = {Interest Groups \& Advocacy},
  year         = {2012},
  volume       = {1},
  pages        = {171--192},
  doi          = {10.1057/iga.2012.9},
  url          = {https://doi.org/10.1057/iga.2012.9}
}

@article{33,
  author       = {Gruetzemacher, Ross and Pilditch, Toby D. and Liang, Hao and Manning, Cameron and Gates, Victoria and Moss, Daniel and Elsey, James W. B. and Sleegers, Willem W. A. and Kilian, Kevin},
  title        = {Implications for governance in public perceptions of societal-scale AI risks},
  journal      = {arXiv},
  year         = {2024},
  doi          = {10.48550/arXiv.2406.06199},
  url          = {https://doi.org/10.48550/arXiv.2406.06199}
}

@misc{34,
  author       = {Hashim, Sophia},
  title        = {AI embraces crypto’s dirty politics},
  year         = {2025},
  month        = {August 26},
  howpublished = {Blog post},
  publisher    = {Transformer},
  note         = {Retrieved September 8, 2025 from \url{https://www.transformernews.ai/p/ai-super-pac-leading-the-future-crypto}}
}

@misc{35,
  author       = {Henderson, Caspar},
  title        = {Superintelligence by Nick Bostrom and A rough ride to the future by James Lovelock -- review},
  year         = {2014},
  month        = {July 17},
  howpublished = {Newspaper article},
  publisher    = {The Guardian},
  note         = {Retrieved September 12, 2025 from \url{https://www.theguardian.com/books/2014/jul/17/superintelligence-nick-brostrom-rough-ride-future-james-lovelock-review}}
}

@book{36,
  author       = {Katz, Elihu and Lazarsfeld, Paul F. and Roper, Elmo},
  title        = {Personal influence: The part played by people in the flow of mass communications},
  publisher    = {Free Press},
  year         = {1955}
}

@misc{37,
  author       = {Kokotajlo, Daniel and Alexander, Scott and Larsen, Tom and Lifland, Eli and Dean, Tom},
  title        = {AI 2027},
  year         = {2025},
  note         = {Retrieved September 12, 2025 from \url{https://ai-2027.com/}}
}

@article{38,
  author       = {Landgraf, Philipp and Stamatogiannakis, Antonios and Yang, Huan},
  title        = {How mortality salience hurts brands with different personalities},
  journal      = {International Journal of Research in Marketing},
  year         = {2024},
  volume       = {41},
  number       = {2},
  pages        = {308--324},
  doi          = {10.1016/j.ijresmar.2023.11.002},
  url          = {https://doi.org/10.1016/j.ijresmar.2023.11.002}
}

@book{39,
  author       = {Lazarsfeld, Paul F. and Berelson, Bernard and Gaudet, Hazel},
  title        = {The people's choice},
  publisher    = {Duell, Sloan \& Pearce},
  year         = {1944}
}

@misc{40,
  author       = {Leicht, A.},
  year         = {2025},
  title        = {Don’t build an {AI} safety movement},
  howpublished = {Blog post on Threading the Needle},
  note         = {Retrieved September 17, 2025, from \url{https://writing.antonleicht.me/p/dont-build-an-ai-safety-movement}}
}

@misc{41,
  author       = {Lima-Strong, C.},
  year         = {2025},
  title        = {Unpacking {Trump’s} {AI} action plan: Cutting rules and speeding roll-out},
  howpublished = {Blog post on Tech Policy},
  note         = {Retrieved September 11, 2025, from \url{https://www.techpolicy.press/unpacking-trumps-ai-action-plan-gutting-rules-and-speeding-rollout}}
}

@article{42,
  author       = {Macdonell, A.},
  year         = {1898},
  title        = {The origin and early history of chess},
  journal      = {Journal of the Royal Asiatic Society},
  volume       = {30},
  number       = {1},
  pages        = {117--141},
  doi          = {10.1017/S0035869X00146246}
}

@article{43,
  author       = {Musa, S.},
  year         = {2016},
  title        = {The impact of {NRA} on the {American} policy},
  journal      = {Journal of Political Sciences \& Public Affairs},
  volume       = {4},
  number       = {4},
  url          = {https://www.researchgate.net/publication/312518552_The_Impact_of_NRA_on_the_American_Policy}
}

@misc{44,
  author       = {{National Association of Attorneys General}},
  year         = {2025},
  title        = {Letter to {AI} companies},
  note         = {Retrieved September 19, 2025, from \url{https://illinoisattorneygeneral.gov/News-Room/Current-News/AI%20Chatbot_FINAL%20(44).pdf?language_id=1}}
}

@article{45,
  author       = {Nolan, J. M.},
  year         = {2017},
  title        = {{“An inconvenient truth”} increases knowledge, concern, and willingness to reduce greenhouse gases},
  journal      = {Environment and Behavior},
  volume       = {42},
  number       = {5},
  pages        = {643--658},
  doi          = {10.1177/0013916509357696},
  note         = {(Original work published 2010)}
}

@article{46,
  author       = {Norris, P. and Curtice, J.},
  year         = {2008},
  title        = {Getting the message out: A two-step model of the role of the internet in campaign communication flows during the 2005 {British} general election},
  journal      = {Journal of Information Technology \& Politics},
  volume       = {4},
  number       = {4},
  pages        = {3--13},
  doi          = {10.1080/19331680801975359}
}

@article{47,
  author       = {Pabian, S. and Hudders, L. and Poels, K. and Stoffelen, F. and De Backer, C. J. S.},
  year         = {2020},
  title        = {Ninety minutes to reduce one’s intention to eat meat: A preliminary experimental investigation on the effect of watching the {Cowspiracy} documentary on intention to reduce meat consumption},
  journal      = {Frontiers in Communication},
  volume       = {5},
  pages        = {Article 69},
  doi          = {10.3389/fcomm.2020.00069}
}

@article{48,
  author       = {Paul, H. L. and Fitzgerald, J.},
  year         = {2021},
  title        = {The dynamics of issue salience: Immigration and public opinion},
  journal      = {Polity},
  volume       = {53},
  number       = {3},
  note         = {July 2021},
  doi          = {10.1086/714144}
}

@article{49,
  author       = {Petruzzelli, A. M. and Mora, L. and Natalicchio, A. and Platania, F. and Hernandez, C. T.},
  year         = {2024},
  title        = {Consumers’ reaction to sci-fi as a source of information for technological development: An empirical analysis},
  journal      = {Technovation},
  volume       = {132},
  pages        = {Article 102970},
  doi          = {10.1016/j.technovation.2024.102970}
}

@misc{50,
  author       = {{Pew Research Center}},
  year         = {2025},
  title        = {U.S. workers are more worried than hopeful about future {AI} use in the workplace},
  howpublished = {Report},
  note         = {Published February 25, 2025. Retrieved September 29, 2025, from \url{https://www.pewresearch.org/social-trends/2025/02/25/u-s-workers-are-more-worried-than-hopeful-about-future-ai-use-in-the-workplace/}}
}

@misc{51,
  author       = {{Reuters} and {Ipsos}},
  year         = {2025},
  title        = {Americans fear {AI} permanently displacing workers, {Reuters}/{Ipsos} poll finds},
  howpublished = {Reuters article},
  note         = {Published August 19, 2025. Retrieved September 29, 2025.}
}

@article{52,
  author       = {Rogers, T. B. and Kuiper, N. A. and Kirker, W. S.},
  year         = {1977},
  title        = {Self-reference and the encoding of personal information},
  journal      = {Journal of Personality and Social Psychology},
  volume       = {35},
  number       = {9},
  pages        = {677--688},
  doi          = {10.1037/0022-3514.35.9.677}
}

@article{53,
  author       = {Rosenblatt, A. and Greenberg, J. and Solomon, S. and Pyszczynski, T. and Lyon, D.},
  year         = {1989},
  title        = {Evidence for terror management theory: {I.} The effects of mortality salience on reactions to those who violate or uphold cultural values},
  journal      = {Journal of Personality and Social Psychology},
  volume       = {57},
  number       = {4},
  pages        = {681--690},
  doi          = {10.1037/0022-3514.57.4.681}
}

@phdthesis{54,
  author       = {Sadeghi, S.},
  year         = {2025},
  title        = {Employee well-being in the age of {AI}: Perceptions, concerns, behaviors, and outcomes},
  school       = {University of the Incarnate Word},
  doi          = {10.48550/arXiv.2412.04796}
}

@article{55,
  author       = {Schofield, J. and Pavelchak, M.},
  year         = {1989},
  title        = {Fallout from \emph{The Day After}: The impact of a {TV} film on attitudes related to nuclear war},
  journal      = {Journal of Applied Social Psychology},
  volume       = {19},
  number       = {4},
  pages        = {433--448},
  doi          = {10.1111/j.1559-1816.1989.tb00066.x}
}

@article{56,
  author       = {Schonger, M. and Sele, D.},
  year         = {2021},
  title        = {Intuition and exponential growth: Bias and the roles of parameterization and complexity},
  journal      = {Mathematische Semesterberichte},
  volume       = {68},
  number       = {2},
  pages        = {221--235},
  doi          = {10.1007/s00591-021-00306-7}
}

@misc{57,
  author       = {{Seismic Foundation}},
  year         = {2025},
  title        = {On the razor’s edge: {AI} vs. everything we care about},
  howpublished = {Report},
  note         = {Retrieved September 8, 2025, from \url{https://report2025.seismic.org/}}
}

@article{58,
  author       = {Siciński, A.},
  year         = {1963},
  title        = {A two-step flow of communication: Verification of a hypothesis in {Poland}},
  journal      = {The Polish Sociological Bulletin},
  volume       = {7},
  pages        = {33--40},
  url          = {http://www.jstor.org/stable/44815133}
}

@article{59,
  author       = {Soffer, O.},
  year         = {2021},
  title        = {Algorithmic personalization and the two-step flow of communication},
  journal      = {Communication Theory},
  volume       = {31},
  number       = {3},
  pages        = {297--315},
  doi          = {10.1093/ct/qtz008}
}

@article{60,
  author       = {Symons, C. S. and Johnson, B. T.},
  year         = {1997},
  title        = {The self-reference effect in memory: A meta-analysis},
  journal      = {Psychological Bulletin},
  volume       = {121},
  number       = {3},
  pages        = {371--394},
  doi          = {10.1037/0033-2909.121.3.371}
}

@misc{61,
  author       = {Taylor, J. and Hern, A.},
  year         = {2023},
  title        = {‘Godfather of {AI}’ {Geoffrey Hinton} quits {Google} and warns over dangers of misinformation},
  howpublished = {The Guardian},
  note         = {Published May 2, 2023. Retrieved September 12, 2025, from \url{https://www.theguardian.com/technology/2023/may/02/geoffrey-hinton-godfather-of-ai-quits-google-warns-dangers-of-machine-learning}}
}

@article{62,
  author       = {Tversky, A. and Kahneman, D.},
  year         = {1973},
  title        = {Availability: A heuristic for judging frequency and probability},
  journal      = {Cognitive Psychology},
  volume       = {5},
  number       = {2},
  pages        = {207--232},
  doi          = {10.1016/0010-0285(73)90033-9}
}

@article{63,
  author       = {Wagenaar, W. A. and Sagaria, S. D.},
  year         = {1975},
  title        = {Misperception of exponential growth},
  journal      = {Perception \& Psychophysics},
  volume       = {18},
  pages        = {416--422},
  doi          = {10.3758/BF03204114}
}

@misc{64,
  author       = {Wang, W. and Toscano, M.},
  year         = {2024},
  title        = {Artificial intelligence and relationships: 1 in 4 young adults believe {AI} partners could replace real-life romance},
  howpublished = {Institute for Family Studies},
  note         = {Published November 14, 2024.}
}

@article{65,
  author       = {Wasow, O.},
  year         = {2020},
  title        = {Agenda seeding: How 1960s {Black} protests moved elites, public opinion, and voting},
  journal      = {American Political Science Review},
  volume       = {114},
  number       = {3},
  pages        = {638--659},
  doi          = {10.1017/S000305542000009X}
}

@article{66,
  author       = {Wlezien, C.},
  year         = {1995},
  title        = {The public as thermostat: Dynamics of preferences for spending},
  journal      = {American Journal of Political Science},
  volume       = {39},
  number       = {4},
  pages        = {981--1000},
  doi          = {10.2307/2111666}
}

@misc{67,
  author       = {Yudkowsky, E.},
  year         = {2007},
  title        = {Cognitive biases potentially affecting judgment of global risks},
  howpublished = {Machine Intelligence Research Institute},
  note         = {Retrieved September 17, 2025, from \url{https://intelligence.org/files/CognitiveBiases.pdf}}
}

@techreport{68,
  author       = {Zhang, B. and Dafoe, A.},
  year         = {2019},
  title        = {Artificial intelligence: {American} attitudes and trends},
  institution  = {SSRN},
  doi          = {10.2139/ssrn.3312874},
  note         = {Published January 9, 2019}
}

@article{69,
  author       = {Zhang, Y. and Zhao, D. and Hancock, J. T. and Kraut, R. and Yang, D.},
  year         = {2025},
  title        = {The rise of {AI} companions: How human-chatbot relationships influence well-being},
  journal      = {arXiv},
  note         = {(arXiv Preprint No. 2506.12605)},
  doi          = {10.48550/arXiv.2506.12605}
}

@misc{70,
  author       = {Ray, J.},
  title        = {Americans express real concerns about artificial intelligence},
  howpublished = {\emph{Gallup}.},
  month        = aug,
  day          = {26},
  year         = {2024},
  note         = {Survey conducted for the Bentley University–Gallup Business in Society Report},
  url          = {https://news.gallup.com/poll/648953/americans-express-real-concerns-artificial-intelligence.aspx}
}

@article{71,
  author    = {Loewenstein, George F. and Weber, Elke U. and Hsee, Christopher K. and Welch, Ned},
  title     = {Risk as feelings},
  journal   = {Psychological Bulletin},
  year      = {2001},
  volume    = {127},
  number    = {2},
  pages     = {267--286},
  doi       = {10.1037/0033-2909.127.2.267},
  url       = {https://doi.org/10.1037/0033-2909.127.2.267}
}

@incollection{72,
  author    = {Wlezien, Christopher and Soroka, Stuart N.},
  title     = {Public opinion and public policy},
  booktitle = {Oxford Research Encyclopedia of Politics},
  publisher = {Oxford University Press},
  year      = {2021},
  doi       = {10.1093/acrefore/9780190228637.013.74},
  url       = {https://doi.org/10.1093/acrefore/9780190228637.013.74}
}

\newpage
\section*{Figures}

\end{document}